\documentclass[aps,prc,reprint,superscriptaddress,nofootinbib,amsmath,amssymb]{revtex4-2}
\usepackage{float}
\usepackage{booktabs}
\usepackage{graphicx} 
\usepackage{bm}       
\usepackage{hyperref} 
\hypersetup{colorlinks=true, citecolor=blue, urlcolor=blue, linkcolor=blue}

\begin{document}

\title{Impact on Final-State Bulk Observables in Heavy-Ion Collisions from an Extended MC Glauber Model of Quantum Rotors}

\author{Yanru Bao}
\email{yanrubao@mails.ccnu.edu.cn}
\affiliation{Key Laboratory of Quark and Lepton Physics (MOE) \& Institute of Particle Physics, Central China Normal University, Wuhan, Hubei 430079, China}

\author{Weiyao Ke}
\email{weiyaoke@ccnu.edu.cn}
\affiliation{Key Laboratory of Quark and Lepton Physics (MOE) \& Institute of Particle Physics, Central China Normal University, Wuhan, Hubei 430079, China}
\affiliation{Southern Center for Nuclear-Science Theory (SCNT), Institute of Modern Physics, Chinese Academy of Sciences, Huizhou, Guangdong 516000, China}
\date{\today}

\begin{abstract}
Nuclear structure is closely connected to anisotropic flow in central heavy-ion collisions (HICs), so flow observables provide important probes for constraining nuclear deformation. Conventional Monte Carlo (MC) models typically describe deformed nuclei as classical rotors with definite orientations, whereas the ground states of even--even nuclei with $J=0$ are rotationally invariant. Motivated by the importance of rotational symmetry restoration, we systematically investigate symmetric collisions involving the representative even--even nuclei $^{20}\mathrm{Ne}$, $^{28}\mathrm{Si}$, $^{96}\mathrm{Ru}$, and $^{238}\mathrm{U}$ using an extended MC Glauber model that incorporates quantum superposition of nuclear orientations. In the final state, the quantum treatment suppresses elliptic flow $v_2\{2\}$, with the strongest effect observed in ultra-central Ne+Ne collisions. Quantum superposition further induces system-dependent changes in normalized symmetric cumulants and shifts the Pearson correlation between $v_2^2$ and $[p_T]$ toward the corresponding near-spherical reference values. Moreover, the quantum treatment modifies cross-system flow ratios and the U+U/Au+Au covariance ratio used experimentally to constrain nuclear deformation parameters. Our results suggest that the quantum-rotor treatment may affect the interpretation of nuclear structure information inffered from heavy-ion collision final-state observables.
\end{abstract}

\maketitle

\section{Introduction}

HICs are used to create quark--gluon plasma (QGP) matter at extremely high temperatures and energy densities. Owing to its extremely short lifetime of about $5$--$10\,\mathrm{fm}/c$, the QGP cannot be observed directly, and its properties are instead inferred by comparing experimental measurements with dynamical model calculations~\cite{Busza:2018rrf,He:2022ywp}. Over the past two decades, observables such as transverse-momentum spectra and anisotropic flow $v_n$ have been extensively used to constrain the transport properties of the QGP\cite{PHENIX:2003iij,Gale:2012rq,Zhao:2017yhj,Nielsen:2022jms}, in particular its shear and bulk viscosities~\cite{Heinz:2013th,JETSCAPE:2020shq,JETSCAPE:2020mzn,Parkkila:2021yha}. More recently, with the increasing precision of both experimental measurements and event-by-event simulations, the same data--model comparison strategy has also been used to extract nuclear-structure parameters~\cite{Giacalone:2019pca,
Giacalone:2020awm,
Jia:2021tzt,
Giacalone:2021udy,
Jia:2021qyu,
Jia:2021wbq,
ATLAS:2022dov,STAR:2024wgy}. For example, the STAR Collaboration measured $\langle v_2^2\rangle$, $\langle(\delta p_T)^2\rangle$, and $\langle v_2^2\delta p_T\rangle$ in U+U and Au+Au collisions, and compared their cross-system ratios with event-by-event hydrodynamic calculations. By varying the deformation parameters in the simulations, these observables can be used to constrain the quadrupole deformation and triaxiality of $^{238}\mathrm{U}$~\cite{STAR:2024wgy}.

A relativistic heavy-ion collision evolves through several stages, including the initial collision and energy deposition, pre-equilibrium evolution, hydrodynamic expansion of the QGP, hadronization, and the subsequent hadronic stage~\cite{Bernhard:2018hnz,Shen:2014vra}. During the evolution, pressure gradients generated by the initial geometry drive the collective expansion and convert spatial anisotropy into momentum-space anisotropy of the final-state particles. Consequently, information about the initial geometry is retained in final-state observables such as anisotropic flow and multiparticle correlations~\cite{Noronha-Hostler:2015dbi,ALICE:2025luc,LHCb:2025ixz,ATLAS:2025nnt,CMS:2025tga,Giacalone:2024luz,ALICE:2016ccg,ATLAS:2012at,Voloshin:2008dg}. Therefore, a reliable description of the initial state is essential for interpreting final-state observables and extracting the underlying physics of the collision~\cite{Shou:2014eya}.

Traditional MC models are commonly used to generate nuclear configurations for heavy-ion collision simulations by treating a deformed nucleus as a classical rotor: in each event, a definite intrinsic orientation is randomly assigned, and the nucleon positions are sampled from the corresponding deformed one-body density~\cite{Miller:2007ri,Moreland:2014oya,Dobaczewski:2025rdi,dEnterria:2020dwq}. For even--even nuclei, however, the physical ground state has $J=0$ and is rotationally invariant. The classical configurations generated by conventional MC models carry definite orientations, which is inconsistent with the rotational invariance of an even--even $J=0$ ground state~\cite{Dobaczewski:2025rdi}. Motivated by the importance of rotational-symmetry restoration, we recently proposed a quantum-mechanical description of nuclear configurations by extending the conventional MC sampling procedure~\cite{Ke:2025tyv}. In this approach, the $J=0$ nucleus is treated as a quantum rotor, with its ground state represented as a coherent superposition of intrinsic states with different orientations, thereby approximately restoring the rotational symmetry of the nuclear ground state. Recent studies have also explored the impact of rotational-symmetry restoration on the interpretation of nuclear deformation in ultrarelativistic heavy-ion collisions using different theoretical approaches~\cite{Xu:2026llz,Bofos:2026nmg}.

Replacing the classical rotors with the approximately symmetry-restored quantum rotors changes the spatial distribution of participant nucleons and thus the collision initial geometry. Since information about the initial geometry is propagated to the final state through the subsequent dynamical evolution, an important question is how the quantum-rotor treatment affects experimentally accessible final-state observables after the full collision evolution. In this work, we systematically investigate this question in symmetric collisions of the representative even--even nuclei $^{20}\mathrm{Ne}$, $^{28}\mathrm{Si}$, $^{96}\mathrm{Ru}$, and $^{238}\mathrm{U}$, covering both light and heavy nuclear systems and prolate and oblate deformations. For each system, we employ both the classical-rotor and quantum-rotor treatments within the same extended MC Glauber framework, while keeping all settings of the subsequent dynamical evolution identical. This allows us to isolate the effects arising from the different treatments of the nuclear configuration. We focus on final-state observables closely related to nuclear deformation, including anisotropic flow~\cite{Voloshin:1994mz}, normalized symmetric cumulants~\cite{ALICE:2016kpq,Gardim:2016nrr,Zhu:2016puf}, the Pearson correlation coefficient between $v_2^2$ and the event-averaged transverse momentum $[p_T]$~\cite{Bozek:2016yoj,Schenke:2020uqq}, and the corresponding cross-system covariance ratio.

The remainder of this paper is organized as follows. Section II describes how nucleons are sampled from classical and quantum rotors using the extended MC Glauber model, and then introduces the subsequent dynamical evolution framework and the observables used in the analysis. Section III presents the effects of quantum-rotor treatment on the initial geometry and final-state observables in different collision systems. Section IV summarizes the main conclusions and discusses possible directions for future studies.

\section{Model setup and observables}

We use TRENTo~\cite{Moreland:2014oya} to construct the initial state of each collision. For each event, the nucleon positions supplied to TRENTo are sampled according to either the classical or the quantum rotor description. Nucleon positions fluctuate from event to event, and their joint distributions carry information about nuclear structure. The nucleon sampling procedures for classical and quantum rotors will be elaborated in the following section. The nucleon--nucleon collisions are then sampled according to the collision criterion, and nucleons that undergo at least one collision are identified as participants. Given the sampled participant nucleons, the TRENTo parametrization of the initial energy deposition is
\begin{equation}
\mathcal E(\mathbf x)
=
\mathcal E_0
\left(
\frac{
n_A^p(\mathbf x)+n_B^p(\mathbf x)
}{2}
\right)^{1/p},
\label{eq:trento_energy_deposition}
\end{equation}
where $n_A(\mathbf x)$ denotes the participant thickness function of nucleus $A$,
\begin{equation}
n_A(\mathbf x)
=
\sum_{i\in A_{\rm part}}
\frac{\gamma_i}{2\pi w^2}
\exp\left[
-\frac{(\mathbf x-\mathbf x_i)^2}{2w^2}
\right],
\label{eq:trento_participant_thickness}
\end{equation}
with $n_B(\mathbf x)$ defined analogously. Here, $\mathbf x_i$ denotes the transverse position of participant nucleon $i$, while $w$ determines the transverse smearing of the energy deposited around each participant nucleon. The weights $\gamma_i$ are independent random variables drawn from a unit-mean Gamma distribution with variance $1/k$, where $k$ controls the magnitude of participant-level fluctuations in the energy deposition. The parameter $p$ determines the generalized-mean ansatz through which the participant thickness functions $n_A$ and $n_B$ are combined, with $p=0$ corresponding to the geometric-mean limit. Finally, $\mathcal E_0$ sets the overall normalization of the initial energy density. In this work, we set $p=0.007$, $k=1.187$, and $w=0.956~\mathrm{fm}$. The same TRENTo parameters and nucleon--nucleon collision prescription are used for the classical and quantum calculations.

The subsequent dynamical evolution is kept identical for the classical and quantum calculations. The TRENTo initial condition first undergoes a short free-streaming pre-hydrodynamic evolution~\cite{Liu:2015nwa} and is then matched to $(2+1)$-dimensional viscous hydrodynamics implemented in VISHNew~\cite{Song:2007ux,Shen:2014vra}. As the system cools, particlization is performed on an isothermal hypersurface using the Cooper--Frye prescription~\cite{Cooper:1974mv,Shen:2014vra}, which converts the hydrodynamic fields into hadrons. These hadrons are then propagated through the UrQMD hadronic cascade~\cite{Bass:1998ca,Bleicher:1999xi}, where they undergo hadronic scatterings and resonance decays. After the hadronic evolution is completed, the final-state observables are calculated from the resulting hadrons.

\subsection{Classical-rotor and orientation-quantum-superposition configurations}

To incorporate the nuclear configurations, TRENTo is coupled to the extended MC Glauber framework developed in Ref.~\cite{Ke:2025tyv}. In this framework, for an axially deformed even--even nucleus $X=A,B$, the rotationally invariant $J=0$ ground state is represented as a coherent superposition of intrinsic configurations with different orientations,
\begin{equation}
|X\rangle
=
\frac{1}{N_X}
\int d\Omega_X\,|\Omega_X\rangle ,
\label{eq:quantum_state}
\end{equation}
where $|\Omega_X\rangle$ denotes an intrinsic configuration with orientation $\Omega_X$, and
\begin{equation}
|N_X|^2
=
\int d\Omega_X d\Omega_X'
\langle\Omega_X|\Omega_X'\rangle .
\label{eq:quantum_norm}
\end{equation}

Within the independent-pair approximation, the corresponding nucleus--nucleus inelastic collision probability for the quantum treatment can be written as~\cite{Ke:2025tyv}
\begin{align}
P_{AB,Q}^{\rm inel}(\mathbf b)
&=
\int d\Omega_A d\Omega_A'
\frac{\langle\Omega_A|\Omega_A'\rangle}{|N_A|^2}
\nonumber\\
&\quad\times
\int d\Omega_B d\Omega_B'
\frac{\langle\Omega_B|\Omega_B'\rangle}{|N_B|^2}
\nonumber\\
&\quad\times
\Bigg\{
1-
\left[
1-
P_{\Omega_A\Omega_A';\Omega_B\Omega_B'}^{NN}(\mathbf b)
\right]^{AB}
\Bigg\},
\label{eq:quantum_glauber_probability}
\end{align}
where the subscript $Q$ denotes the quantum treatment. This expression leads naturally to the Monte Carlo interpretation used in our calculation. For each nucleus $X$, a pair of intrinsic orientations $(\Omega_X,\Omega_X')$ is sampled, with its statistical weight determined by
\begin{equation}
\frac{\langle\Omega_X|\Omega_X'\rangle}{|N_X|^2}.
\label{eq: quantum_weight}
\end{equation}
Unlike the classical-rotor treatment, which samples a single orientation, the quantum-rotor treatment samples a pair of orientations, thereby retaining the off-diagonal information in the collective coordinates.

To construct the nuclear configuration corresponding to a sampled orientation pair, we apply the phase-space prescription developed in our previous work~\cite{Ke:2025tyv}. The phase-space overlap between two intrinsic orientations is approximated by the geometric mean of their phase-space distributions,
\begin{equation}
\mathcal F_X
(\mathbf r,\mathbf p;\Omega_X,\Omega_X')
=
\frac{
\sqrt{
f_X(\mathbf r,\mathbf p;\Omega_X)
f_X(\mathbf r,\mathbf p;\Omega_X')
}
}{
(2\pi)^3
},
\label{eq:phase_space_overlap}
\end{equation}
where $f_X(\mathbf r,\mathbf p;\Omega_X)$ denotes the single-particle phase-space distribution for the intrinsic configuration with orientation $\Omega_X$. It is constrained such that its momentum integral reproduces the corresponding deformed nuclear density,
\begin{equation}
g
\int
\frac{d^3p}{(2\pi)^3}
f_X(\mathbf r,\mathbf p;\Omega_X)
=
\rho_X(\mathbf r;\Omega_X),
\label{eq:phase_space_density}
\end{equation}
where $g$ denotes the spin-isospin degeneracy factor. The nuclear density is parametrized by a deformed Woods--Saxon distribution~\cite{Shou:2014eya},
\begin{equation}
\rho_X(\mathbf r;\Omega_X)
=
\frac{\rho_{0,X}}
{
1+
\exp\left[
\frac{r-R_X(\theta,\phi)}{a_X}
\right]
},
\label{eq:woods_saxon_density}
\end{equation}
where $\rho_{0,X}$ is the central nuclear density, $a_X$ is the surface diffuseness, and $R_X(\theta,\phi)$ is the angle-dependent nuclear radius,
\begin{align}
R_X(\theta,\phi)
&=
R_{0,X}
\Big[
1
+\beta_{2,X}Y_{20}(\theta,\phi)
+\beta_{3,X}Y_{30}(\theta,\phi)
\notag\\
&\qquad
+\beta_{4,X}Y_{40}(\theta,\phi)
+\cdots
\Big].
\label{eq:deformed_radius}
\end{align}
Here, $R_{0,X}$ denotes the reference nuclear radius, $\beta_{\ell,X}$ are the corresponding multipole deformation parameters, and $Y_{\ell 0}(\theta,\phi)$ are the spherical harmonics describing the angular dependence of the axially symmetric multipole deformations. The nuclear structure parameters used in this work are listed in Table~\ref{tab:nuclear_parameters}, with $^{16}\mathrm{O}$ and $^{197}\mathrm{Au}$ approximated as spherical nuclei.

\begin{table}[t]
\centering
\caption{Nuclear parameters for selected isotopes~\cite{Moller:2015fba}.}
\label{tab:nuclear_parameters}
\begin{tabular}{lcccccc}
\hline
Nucleus $X$ & $A$ & $\beta_{2,X}$ & $\beta_{3,X}$ & $\beta_{4,X}$ & $R_{0,X}$ (fm) & $a_X$ (fm) \\
\hline
$^{16}$O  & 16  & 0       & 0      & 0       & 2.608  & 0.513 \\
$^{20}$Ne & 20  & 0.4899  & 0.2160 & 0.3055  & 2.724  & 0.498 \\
$^{28}$Si & 28  & -0.313  & 0      & 0.164   & 3.34   & 0.58  \\
$^{96}$Ru & 96  & 0.16    & 0      & 0       & 5.065  & 0.485 \\
$^{197}$Au& 197 & 0  & 0      & 0  & 6.38   & 0.535 \\
$^{238}$U & 238 & 0.236   & 0      & 0.098   & 6.8054 & 0.605 \\
\hline
\end{tabular}
\end{table}

Within the same extended MC Glauber framework, the classical rotors
are obtained by replacing the quantum orientation-pair weight in
Eq.~\eqref{eq: quantum_weight} with
\begin{equation}
\frac{\langle\Omega_X|\Omega_X'\rangle}{|N_X|^2}
\;\longrightarrow\;
\frac{1}{4\pi}\,
\delta^{(2)}(\Omega_X-\Omega_X').
\label{eq:classical_overlap_limit}
\end{equation}
This replacement restricts the orientation pair to $\Omega_X'=\Omega_X$, so that each retained term corresponds to a definite intrinsic orientation, thereby recovering the classical-rotor picture from the quantum-rotor description.

Correspondingly, the nucleus--nucleus inelastic collision probability
reduces to the classical-rotor form~\cite{Ke:2025tyv},
\begin{align}
P_{AB,C}^{\rm inel}(\mathbf b)
&=
\int\frac{d\Omega_A}{4\pi}
\int\frac{d\Omega_B}{4\pi}
\nonumber\\
&\quad\times
\Bigg\{
1-
\left[
1-
P_{\Omega_A\Omega_A;\Omega_B\Omega_B}^{NN}(\mathbf b)
\right]^{AB}
\Bigg\},
\label{eq:classical_glauber_probability}
\end{align}
where the subscript $C$ denotes the classical treatment. In this limit, the phase-space overlap in Eq.~\eqref{eq:phase_space_overlap} becomes
\begin{equation}
\left.
\mathcal F_X
(\mathbf r,\mathbf p;\Omega_X,\Omega_X')
\right|_{\Omega_X'=\Omega_X}
=
\frac{
f_X(\mathbf r,\mathbf p;\Omega_X)
}{
(2\pi)^3
}.
\label{eq:classical_phase_space_limit}
\end{equation}
Together with Eq.~\eqref{eq:phase_space_density}, its momentum integral reproduces the intrinsic nuclear density,
\begin{equation}
g\int d^3p\,
\mathcal F_X(\mathbf r,\mathbf p;\Omega_X,\Omega_X)
=
\rho_X(\mathbf r;\Omega_X).
\label{eq:classical_spatial_density}
\end{equation}
Thus, the classical prescription corresponds to sampling a single orientation uniformly and generating nucleon positions from the corresponding deformed Woods--Saxon density.

For spherical nuclei, the intrinsic phase-space distribution is independent of orientation,
$f_X(\mathbf r,\mathbf p;\Omega_X)
=
f_X^{\mathrm{sph}}(\mathbf r,\mathbf p)$.
The phase-space overlap therefore reduces to
\begin{equation}
\mathcal F_X(\mathbf r,\mathbf p;\Omega_X,\Omega_X')
=
\frac{f_X^{\mathrm{sph}}(\mathbf r,\mathbf p)}{(2\pi)^3}
\label{eq:spherical_phase_space}
\end{equation}
for any orientation pair, including $\Omega_X'\ne\Omega_X$.
Combining Eq.~\eqref{eq:spherical_phase_space} with
Eq.~\eqref{eq:phase_space_density}, we obtain
\begin{align}
g\int\frac{d^3p}{(2\pi)^3}\,
f_X^{\mathrm{sph}}(\mathbf r,\mathbf p)
=
\rho_X^{\mathrm{sph}}(r),
\label{eq:spherical_spatial_density}
\end{align}
where $\rho_X^{\mathrm{sph}}(r)$ is the Woods--Saxon density with all deformation parameters set to zero. In this spherical limit, the quantum and classical configurations therefore become equivalent within our sampling framework. In the current study, $^{16}\mathrm{O}$ and $^{197}\mathrm{Au}$ are treated as approximately spherical nuclei, and we use only classical rotors as input for O+O and Au+Au collisions in the subsequent calculations.

\subsection{Observables}
In this section, we introduce not only the experimentally accessible final-state observables but also the initial-state eccentricity. Since anisotropic flow responds approximately linearly to the initial eccentricity, the latter provides a way to understand the qualitative trends of quantum effects at a lower computational cost.

\paragraph{Eccentricity}The anisotropic geometry of the initial fireball is quantified by the complex eccentricity moments constructed from the midrapidity energy density $\mathcal{E}(\mathbf{x})$ deposited by TRENTo~\cite{Ke:2025tyv},
\begin{equation}
\epsilon_n
=
\frac{
\int \mathcal{E}(\mathbf{x})\, e^{in\phi}\, |\mathbf{x}|^n\, d^{2}\mathbf{x}
}{
\int \mathcal{E}(\mathbf{x})\, |\mathbf{x}|^n\, d^{2}\mathbf{x}
},
\label{eq:epsilon_n}
\end{equation}
where $\phi$ is the azimuthal angle of the transverse coordinate $\mathbf{x}$. We report the root-mean-square(RMS) eccentricity $\sqrt{\langle|\epsilon_n|^2\rangle}$, which characterizes the typical magnitude of the initial-state geometric anisotropy. For the initial-state analysis, we focus on $n=2$, since the elliptic eccentricity $\epsilon_2$ is particularly sensitive to the deformation and orientation of the colliding nuclei and therefore provides a useful measure of how the treatment of nuclear configurations modifies the initial geometry~\cite{Giacalone:2019pca,Giacalone:2020awm}.

The initial spatial anisotropy is subsequently converted by the hydrodynamic evolution into anisotropic flow in the final state. In particular, the elliptic flow approximately follows a linear response to the initial elliptic eccentricity, $v_2 \simeq \kappa_2 |\epsilon_2|$~\cite{Noronha-Hostler:2015dbi}. This makes $v_2$ a natural final-state observable for tracing modifications of the initial geometry induced by different nuclear rotors.

\paragraph{Anisotropic flow}The anisotropic flow of the produced charged hadrons is characterized by the complex flow vector
\begin{equation}
V_n
\equiv
v_n e^{in\Psi_n},
\label{eq:Vn}
\end{equation}
where $v_n$ is the $n$th-order flow harmonic and $\Psi_n$ is the corresponding flow-plane angle. For an event containing $M$ particles, the $Q$ vector is defined as
\begin{equation}
Q_n
=
\sum_{j=1}^{M} e^{in\phi_j},
\label{eq:Qn}
\end{equation}
where $\phi_j$ is the azimuthal angle of particle $j$.

The flow coefficients are calculated using the $Q$-cumulant method~\cite{Bilandzic:2010jr}, which expresses multiparticle azimuthal correlations in terms of the $Q$ vectors while removing self-correlations. The two-particle azimuthal correlation in a single event is defined as
\begin{equation}
\begin{aligned}
\langle 2\rangle_n
&\equiv
\frac{1}{W_2}
\sum_{i,j=1}^{M}{}^{\prime}
e^{in(\phi_i-\phi_j)}
\\
&=
\frac{|Q_n|^2-M}{W_2},
\end{aligned}
\label{eq:two_particle_corr}
\end{equation}
where
\begin{equation}
W_2
=
M(M-1)
\label{eq:W2}
\end{equation}
is the number of ordered distinct particle pairs in the event. Here and below, the prime on a multiparticle sum indicates that all particle indices are distinct. The two-particle cumulant is obtained by averaging the single-event correlations over events using $W_2$ as the event weight,
\begin{equation}
c_n\{2\}
=
\left\langle\!\left\langle 2\right\rangle\!\right\rangle_n
=
\frac{
\sum_{\mathrm{events}}
W_2\,\langle 2\rangle_n
}{
\sum_{\mathrm{events}} W_2
},
\label{eq:cn2}
\end{equation}
For the multiparticle correlators throughout this subsection, single angular brackets denote averages over distinct particle combinations within one event, whereas double angular brackets denote the corresponding weighted averages over events. The corresponding two-particle flow coefficient is
\begin{equation}
v_n\{2\}
=
\sqrt{c_n\{2\}}.
\label{eq:vn2}
\end{equation}

\paragraph{Normalized symmetric cumulant}Correlations between event-by-event fluctuations of different flow harmonics are characterized by the symmetric cumulant~\cite{ALICE:2016kpq},
\begin{equation}
\mathrm{SC}^{v}(m,n)
=
\langle v_m^2 v_n^2\rangle
-
\langle v_m^2\rangle
\langle v_n^2\rangle,
\qquad m\neq n.
\label{eq:SCmn}
\end{equation}
The symmetric cumulant measures the covariance between the squared magnitudes of two different flow harmonics. Since the magnitude of $\mathrm{SC}^{v}(m,n)$ depends on the magnitudes of the individual flow harmonics, it is useful to remove this overall scale dependence by introducing the normalized symmetric cumulant~\cite{ALICE:2016kpq},
\begin{equation}
\mathrm{NSC}_{v,mn}
=
\frac{
\mathrm{SC}^{v}(m,n)
}{
\langle v_m^2\rangle
\langle v_n^2\rangle
}
=
\frac{
\langle v_m^2 v_n^2\rangle
}{
\langle v_m^2\rangle
\langle v_n^2\rangle
}
-1.
\label{eq:NSmn}
\end{equation}
The normalization reduces the dependence on the overall magnitudes of $v_m$ and $v_n$, facilitating comparisons of the correlation strength between different harmonic combinations and collision systems. A positive value of $\mathrm{NSC}_{v,mn}$ indicates a positive correlation between the event-by-event fluctuations of $v_m^2$ and $v_n^2$, whereas a negative value indicates an anticorrelation. In the absence of correlations between the two harmonics, $\mathrm{NSC}_{v,mn}=0$.

The moments entering Eqs.~\eqref{eq:SCmn} and~\eqref{eq:NSmn} are evaluated using mixed-harmonic multiparticle correlations within the same $Q$-cumulant framework. The two-particle correlations are given by Eq.~\eqref{eq:two_particle_corr}, while the mixed-harmonic four-particle correlation in a single event is defined as~\cite{Bilandzic:2013kga}
\begin{align}
\langle 4\rangle_{m,n}
&\equiv
\frac{1}{W_4}
\sum_{\substack{i,j,k,l=1}}^{M}{}^{\prime}
e^{i(m\phi_i+n\phi_j-m\phi_k-n\phi_l)}
\notag\\
&=
\frac{1}{W_4}
\Big[
|Q_m|^2|Q_n|^2
-2\,\mathrm{Re}
\left(
Q_{m+n}Q_m^{*}Q_n^{*}
\right)
\label{eq:four_particle_mixed}
\\
&\quad
-2\,\mathrm{Re}
\left(
Q_m Q_{m-n}^{*}Q_n^{*}
\right)
+|Q_{m+n}|^2
+|Q_{m-n}|^2
\notag\\
&\quad
-(M-4)
\left(
|Q_m|^2+|Q_n|^2
\right)
+M(M-6)
\Big].
\notag
\end{align}
Here,
\begin{equation}
W_4
=
M(M-1)(M-2)(M-3)
\label{eq:W4}
\end{equation}
The mixed-harmonic four-particle correlation is then averaged over events using $W_4$ as the event weight,
\begin{equation}
\left\langle\!\left\langle 4\right\rangle\!\right\rangle_{m,n}
=
\frac{
\sum_{\mathrm{events}}
W_4\,\langle 4\rangle_{m,n}
}{
\sum_{\mathrm{events}} W_4
}.
\label{eq:four_particle_event_average}
\end{equation}

The symmetric cumulant can therefore be expressed directly in terms of the two- and four-particle correlations as
\begin{equation}
\mathrm{SC}^{v}(m,n)
=
\left\langle\!\left\langle 4\right\rangle\!\right\rangle_{m,n}
-
\left\langle\!\left\langle 2\right\rangle\!\right\rangle_m
\left\langle\!\left\langle 2\right\rangle\!\right\rangle_n,
\label{eq:SC_Qcumulant}
\end{equation}
and the corresponding normalized symmetric cumulant is
\begin{equation}
\mathrm{NSC}_{v,mn}
=
\frac{
\left\langle\!\left\langle 4\right\rangle\!\right\rangle_{m,n}
}{
\left\langle\!\left\langle 2\right\rangle\!\right\rangle_m
\left\langle\!\left\langle 2\right\rangle\!\right\rangle_n
}
-1.
\label{eq:NSC_Qcumulant}
\end{equation}

\paragraph{Pearson correlation coefficient}Finally, we study the correlation between anisotropic flow and the event-averaged transverse momentum. For each event, the mean transverse momentum is defined as
\begin{equation}
[p_T]
=
\frac{1}{M}
\sum_{i=1}^{M} p_{T,i},
\label{eq:mean_pT}
\end{equation}
and the reference mean transverse momentum over the selected event sample is obtained by averaging $[p_T]$ with equal event weight,
\begin{equation}
\left\langle\!\left\langle p_T\right\rangle\!\right\rangle
=
\frac{1}{N_{\mathrm{evt}}}
\sum_{\mathrm{events}} [p_T]
,
\label{eq:global_mean_pT}
\end{equation}
where $N_{\mathrm{evt}}$ is the number of events in the selected event sample. The corresponding transverse-momentum fluctuation is defined as
\begin{equation}
\delta p_T
=
[p_T]
-
\left\langle\!\left\langle p_T\right\rangle\!\right\rangle.
\label{eq:delta_pT}
\end{equation}
The correlation between $v_n^2$ and $[p_T]$ is characterized by the
Pearson correlation coefficient~\cite{Bozek:2016yoj,Bozek:2020drh},
\begin{equation}
\rho_{n,\mathrm{pcc}}
=
\frac{
\mathrm{Cov}(v_n^2,[p_T])
}{
\sqrt{\mathrm{Var}(v_n^2)}
\sqrt{\mathrm{Var}([p_T])}
},
\label{eq:rho_pcc}
\end{equation}
which quantifies the event-by-event correlation between anisotropic flow
and transverse-momentum fluctuations and is sensitive to deformation-induced
changes in the initial nuclear geometry. We also consider the related
normalized correlator~\cite{STAR:2024wgy,STAR:2025elk},
\begin{equation}
\rho_n
=
\frac{
\mathrm{Cov}(v_n^2,[p_T])
}{
\langle v_n^2\rangle
\sqrt{\mathrm{Var}([p_T])}
}.
\label{eq:rho_star}
\end{equation}
Compared with $\rho_{n,\mathrm{pcc}}$, $\rho_n$ replaces
$\sqrt{\mathrm{Var}(v_n^2)}$ by $\langle v_n^2\rangle$, providing a
cleaner spherical baseline and making the deformation-driven contribution
more transparent in nuclear-structure studies.

The covariance and variances appearing in these definitions are
\begin{align}
\mathrm{Cov}(v_n^2,[p_T])
&=
\left\langle
\left(v_n^2-\langle v_n^2\rangle\right)
\delta p_T
\right\rangle
\label{eq:cov_definition}
\\
&=
\langle v_n^2[p_T]\rangle
-
\langle v_n^2\rangle
\left\langle\!\left\langle p_T\right\rangle\!\right\rangle,
\notag
\\
\mathrm{Var}([p_T])
&=
\left\langle(\delta p_T)^2\right\rangle,
\label{eq:var_pT_definition}
\\
\mathrm{Var}(v_n^2)
&=
\left\langle
\left(v_n^2-\langle v_n^2\rangle\right)^2
\right\rangle
\label{eq:var_vn2_definition}
\\
&=
\langle v_n^4\rangle
-
\langle v_n^2\rangle^2.
\notag
\end{align}

For convenience, we define the transverse-momentum moments
\begin{equation}
S_k
=
\sum_{i=1}^{M} p_{T,i}^{\,k},
\qquad k=1,2,
\label{eq:Sk}
\end{equation}
and the $p_T$-weighted vector
\begin{equation}
O_n
=
\sum_{i=1}^{M}
p_{T,i}e^{in\phi_i}.
\label{eq:On}
\end{equation}

In addition to the two- and four-particle flow correlations introduced above, the calculation requires a mixed three-particle correlation between the $n$th-order flow harmonic and transverse momentum,
\begin{align}
\langle 3\rangle_n
&\equiv
\frac{1}{W_3}
\sum_{\substack{i,j,k=1}}^{M}{}^{\prime}
e^{in(\phi_i-\phi_j)}p_{T,k}
\label{eq:pearson_three}
\\
&=
\frac{1}{W_3}
\Big[
|Q_n|^2S_1
-MS_1
-2\,\mathrm{Re}\!\left(O_nQ_n^*\right)
+2S_1
\Big],
\notag
\end{align}
where
\begin{equation}
W_3
=
M(M-1)(M-2)
\label{eq:pearson_W3}
\end{equation}
is the number of ordered distinct particle triplets.

The transverse-momentum fluctuations are evaluated through the two-particle $p_T$ correlation,
\begin{equation}
\langle 2_{p_T}\rangle
\equiv
\frac{1}{W_2}
\sum_{i\neq j}p_{T,i}p_{T,j}
=
\frac{S_1^2-S_2}{W_2},
\label{eq:pearson_two_pt}
\end{equation}
where $W_2$ is defined in Eq.~\eqref{eq:W2}. The corresponding event-averaged correlations are
\begin{align}
\left\langle\!\left\langle3\right\rangle\!\right\rangle_n
&=
\frac{
\sum_{\mathrm{events}}
W_3\,\langle3\rangle_n
}{
\sum_{\mathrm{events}}W_3
},
\label{eq:pearson_three_event_average}
\\
\left\langle\!\left\langle2_{p_T}\right\rangle\!\right\rangle
&=
\frac{
\sum_{\mathrm{events}}
W_2\,\langle2_{p_T}\rangle
}{
\sum_{\mathrm{events}}W_2
}.
\label{eq:pearson_two_pt_event_average}
\end{align}
The two-particle flow correlation and its event average are given by Eqs.~\eqref{eq:two_particle_corr} and~\eqref{eq:cn2}, respectively. The same-harmonic four-particle correlation $\langle4\rangle_n$ and its event average $\left\langle\!\left\langle4\right\rangle\!\right\rangle_n$, required below, are obtained from Eqs.~\eqref{eq:four_particle_mixed} and~\eqref{eq:four_particle_event_average} by setting $m=n$.

With these multiparticle estimators, the quantities in Eqs.~\eqref{eq:cov_definition}--\eqref{eq:var_vn2_definition} are evaluated as
\begin{align}
\mathrm{Cov}(v_n^2,[p_T])
&=
\left\langle\!\left\langle3\right\rangle\!\right\rangle_n
-
\left\langle\!\left\langle2\right\rangle\!\right\rangle_n
\left\langle\!\left\langle p_T\right\rangle\!\right\rangle,
\label{eq:cov_vn2_pT}
\\
\mathrm{Var}([p_T])
&=
\left\langle\!\left\langle2_{p_T}\right\rangle\!\right\rangle
-
\left\langle\!\left\langle p_T\right\rangle\!\right\rangle^2,
\label{eq:var_pT}
\\
\mathrm{Var}(v_n^2)
&=
\left\langle\!\left\langle4\right\rangle\!\right\rangle_n
-
\left(
\left\langle\!\left\langle2\right\rangle\!\right\rangle_n
\right)^2.
\label{eq:var_vn2}
\end{align}

Accordingly, the two observables are evaluated as
\begin{align}
\rho_{n,\mathrm{pcc}}
&=
\frac{
\left\langle\!\left\langle3\right\rangle\!\right\rangle_n
-
\left\langle\!\left\langle2\right\rangle\!\right\rangle_n
\left\langle\!\left\langle p_T\right\rangle\!\right\rangle
}{
\sqrt{
\mathrm{Var}(v_n^2)\,
\mathrm{Var}([p_T])
}
},
\label{eq:rho_pcc_final}
\\
\rho_n
&=
\frac{
\left\langle\!\left\langle3\right\rangle\!\right\rangle_n
-
\left\langle\!\left\langle2\right\rangle\!\right\rangle_n
\left\langle\!\left\langle p_T\right\rangle\!\right\rangle
}{
\left\langle\!\left\langle2\right\rangle\!\right\rangle_n
\sqrt{\mathrm{Var}([p_T])}
}.
\label{eq:rho_star_final}
\end{align}

Unless otherwise specified, the final-state observables defined above are calculated using charged hadrons within $|\eta|<0.8$. For the light-ion systems Ne+Ne, Si+Si, and O+O at $\sqrt{s_{NN}}=5.36~\mathrm{TeV}$, the transverse-momentum range is $0.5<p_T<5~\mathrm{GeV}/c$, while for the heavy-ion systems Ru+Ru, U+U, and Au+Au at $\sqrt{s_{NN}}=200~\mathrm{GeV}$, we use $0.2<p_T<3~\mathrm{GeV}/c$.

\section{Results and Discussion}

\subsection{The mass number and shape deformation dependence of the quantum rotor correction to eccentricity}

\begin{figure}[t]
\centering
\includegraphics[width=\columnwidth]{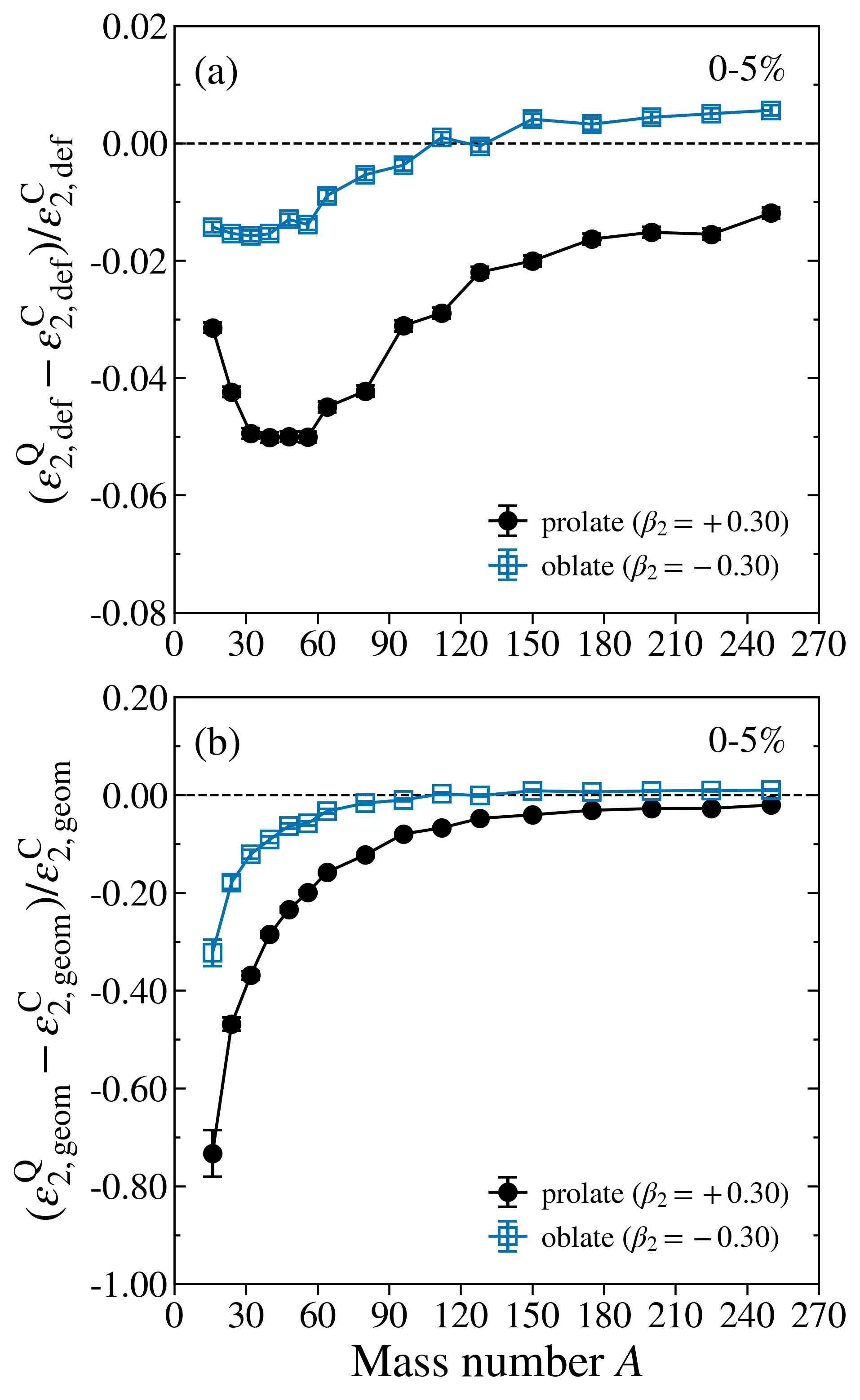}
\caption{
Relative quantum corrections $(Q-C)/C$ to
$\varepsilon_{2,\mathrm{def}}$ (a) and
$\varepsilon_{2,\mathrm{geom}}$ (b)
as functions of the mass number $A$ in $0$--$5\%$ most central collisions
at $\sqrt{s_{NN}}=5.36$~TeV.
}
\label{fig:eps2_geom}
\end{figure}

Before discussing final-state observables, we first investigate the mass-number and shape dependence of the quantum effect in the initial state. For nuclei $X_A^{\rm prol}$ and $X_A^{\rm obl}$, we set
the quadrupole deformation parameters to
$\beta_2=+0.30$ and $\beta_2=-0.30$, respectively, while keeping $\beta_3=\beta_4=0$ for both. Their Woods--Saxon density distributions take the form in Eqs.~\eqref{eq:woods_saxon_density} and~\eqref{eq:deformed_radius}. The radius parameter is determined by $R_0=1.2A^{1/3}$~fm, while the surface diffuseness is fixed at $a=0.52$~fm. In the following, we use $X_A^{\rm def}$ to denote
either $X_A^{\rm prol}$ or $X_A^{\rm obl}$. We consider mass numbers
\begin{equation*}
\begin{aligned}
A={}&16,\,24,\,32,\,40,\,48,\,56,\,64,\,80,\\
   &96,\,112,\,128,\,150,\,175,\,200,\,225,\,250.
\end{aligned}
\end{equation*}
For each value of $A$, we simulate symmetric $X_A^{\rm{def}}+X_A^{\rm{def}}$ collisions at $\sqrt{s_{NN}}=5.36$~TeV using both the classical and quantum nuclear configurations. In the $0$--$5\%$ most central collisions, the corresponding RMS elliptic eccentricity is calculated as
$\varepsilon_{2,\mathrm{def}}
=
\sqrt{
\langle
|\mathcal{E}_{2,\mathrm{def}}|^2
\rangle
}$. 

To quantify the quantum modification and compare it systematically across different mass numbers, we use the relative difference
$(\varepsilon_{2,\mathrm{def}}^Q-\varepsilon_{2,\mathrm{def}}^C)/
\varepsilon_{2,\mathrm{def}}^C$, where the superscripts $Q$ and $C$ denote the quantum and classical rotors, respectively. As shown in Fig.~\ref{fig:eps2_geom}(a), the relative quantum correction exhibits a nonmonotonic dependence on the mass number, with the magnitude of the quantum suppression initially increasing and then decreasing as $A$ increases. Notably, the quantum correction remains appreciable even in the heavy-mass region, indicating that quantum superposition effects should also be considered in heavy-nucleus collisions. 
Since elliptic flow approximately follows a linear response to the initial elliptic eccentricity, $v_2\simeq\kappa_2\varepsilon_2$, these results suggest that quantum superposition effects should be considered in final-state observables for both light- and heavy-nucleus collisions.

At first glance, the nonmonotonic mass-number dependence shown in Fig.~\ref{fig:eps2_geom}(a) appears to differ from the expectation based on the orientation-overlap picture of the quantum-rotor framework~\cite{Ke:2025tyv}.
As the mass number increases, the overlap between intrinsic configurations with different orientations becomes increasingly localized around $\Omega_X\simeq\Omega_X'$, gradually recovering the classical-rotor limit.
Consequently, the quantum correction to the deformation-induced geometry is expected to decrease monotonically in magnitude with increasing $A$. Although orientation quantum superposition primarily modifies the deformation-induced geometry, the elliptic eccentricity is influenced by both geometric contributions and fluctuations. The mean-squared elliptic eccentricity can be approximately parameterized as~\cite{Jia:2021tzt}
\begin{equation}
\left\langle |\mathcal{E}_2|^2 \right\rangle
\simeq
a_2'
+
\left[
b_2'+c_2'\operatorname{sgn}(\beta_2)
\right]\beta_2^2,
\label{eq:eps2_deformation_complex}
\end{equation}
where $a_2'$ represents the spherical fluctuation baseline, while the second term describes the deformation-dependent contribution.

To isolate the geometric contribution and examine whether the expected monotonic mass-number dependence can be recovered, we construct a corresponding spherical nucleus $X_A^{\rm sph}$ for each deformed nucleus $X_A^{\rm def}$, keeping all parameters unchanged except for setting $\beta_2=0$. For the deformed system, Eq.~\eqref{eq:eps2_deformation_complex} gives
\begin{equation}
\left\langle |\mathcal{E}_{2,\mathrm{def}}|^2 \right\rangle
\simeq
a_2' + \left[b_2'+c_2'\operatorname{sgn}(\beta_2)\right]\beta_2^2,
\label{eq:eps2_def_new}
\end{equation}
whereas for the corresponding spherical system,
\begin{equation}
\left\langle |\mathcal{E}_{2,\mathrm{sph}}|^2 \right\rangle
\simeq a_2'.
\label{eq:eps2_sph_new}
\end{equation}
The spherical calculation therefore provides a direct estimate of the baseline at the same mass number and under the same collision conditions. Subtracting Eq.~\eqref{eq:eps2_sph_new} from Eq.~\eqref{eq:eps2_def_new} gives
\begin{equation}
\left\langle |\mathcal{E}_{2,\mathrm{def}}|^2 \right\rangle
-\left\langle |\mathcal{E}_{2,\mathrm{sph}}|^2 \right\rangle \simeq
\left[b_2'+c_2'\operatorname{sgn}(\beta_2)\right]\beta_2^2,
\label{eq:eps2_deformation_part}
\end{equation}
which approximately isolates the deformation-dependent contribution at the level of the mean-squared eccentricity. To express this contribution in the same RMS form as the eccentricity itself, we define
\begin{equation}
\varepsilon_{2,\mathrm{geom}}
\equiv
\sqrt{
\left\langle |\mathcal{E}_{2,\mathrm{def}}|^2 \right\rangle
-
\left\langle |\mathcal{E}_{2,\mathrm{sph}}|^2 \right\rangle
}.
\label{eq:eps2_geom}
\end{equation}
This quantity provides a deformation-sensitive measure of the geometric contribution to the elliptic eccentricity.

The corresponding results are shown in Fig.~\ref{fig:eps2_geom}(b). After subtracting the spherical baseline, the negative quantum corrections to $\varepsilon_{2,\mathrm{geom}}$ become progressively weaker with increasing $A$ for both prolate and oblate nuclei, approaching zero in the heavy-mass region. This behavior is consistent with the orientation-overlap picture of Ref.~\cite{Ke:2025tyv}, indicating that the quantum modification of the deformation-induced geometry generally weakens with increasing mass number.

\begin{figure}[t]
\centering
\includegraphics[width=\columnwidth]{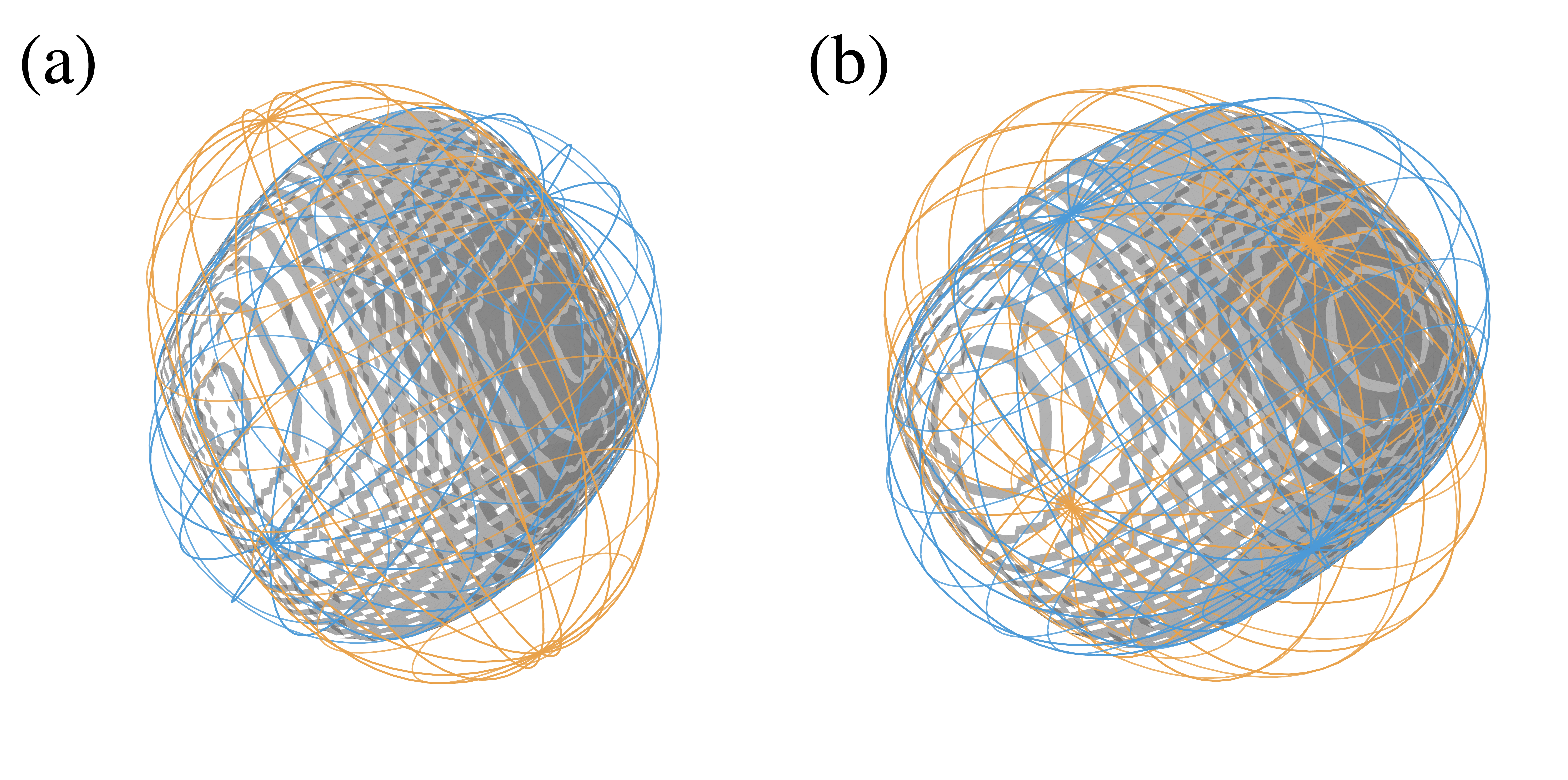}
\caption{Schematic illustrations of the spatial overlap between two intrinsic orientations for (a) prolate and (b) oblate nuclei. In each panel, the blue and orange wireframes correspond to two orientations, $\Omega_X$ and $\Omega_X'$, while the gray region illustrates their spatial overlap. The relative angle between the two orientations is fixed at $\Delta\Omega=\pi/2$ in both panels.}
\label{fig:orientation_overlap}
\end{figure}

The nonmonotonic mass-number dependence observed for the total eccentricity in Fig.~\ref{fig:eps2_geom}(a) can therefore be understood from the competition between geometric quantum effects and finite-nucleon fluctuations. For light nuclei, the geometric quantum modification is relatively strong, but its contribution to the total eccentricity is diluted by large fluctuations. As the mass number increases, the fluctuation contribution becomes less important, making the geometric quantum effect more visible. 

Furthermore, Fig.~\ref{fig:eps2_geom} reveals a pronounced difference between prolate and oblate nuclei. At the same magnitude of quadrupole deformation, $|\beta_2|=0.30$, the quantum correction is generally much stronger for prolate nuclei than for oblate nuclei. This shape dependence persists after subtracting the spherical fluctuation baseline, as shown in Fig.~\ref{fig:eps2_geom}(b). These results demonstrate that the quantum modification of the initial geometry depends not only on the mass number and deformation magnitude, but also on the sign of $\beta_2$.

This difference can be qualitatively understood from the spatial overlap between intrinsic configurations with different orientations, as illustrated schematically in Fig.~\ref{fig:orientation_overlap}. For two nearly orthogonal prolate configurations, the overlap between their rod-like densities is mainly concentrated near the nuclear center, producing a blob-like region. This geometric picture can be summarized schematically as ``rod $\times$ rod $\rightarrow$ blob-like overlap region.'' In contrast, two nearly orthogonal oblate configurations have disk-like densities whose overlap extends along a common diameter, forming a rod-like region with prolate-like geometric anisotropy. The corresponding geometric picture can be summarized schematically as ``disk $\times$ disk $\rightarrow$ rod-like overlap region.'' Consequently, while the overlap of prolate configurations tends to suppress the geometric anisotropy, that of oblate configurations can retain a substantial degree of anisotropy even after orientation superposition. This geometric difference provides a possible explanation for the weaker quantum suppression observed in oblate nuclei compared with prolate nuclei.

\subsection{Quantum effects on final-state observables}

\begin{figure}[t]
\centering
\includegraphics[width=\columnwidth]{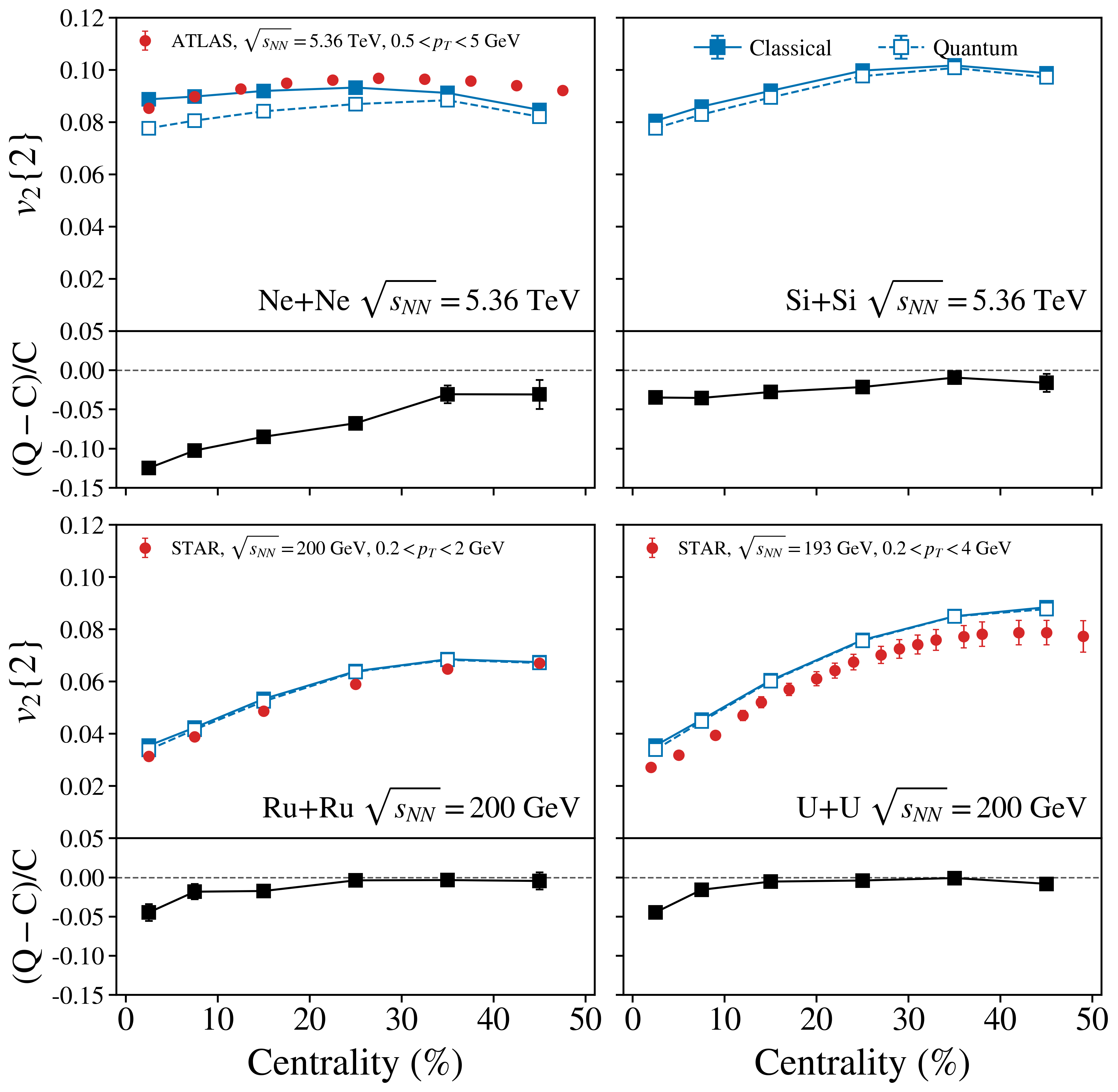}
\caption{Centrality dependence of $v_2\{2\}$ in Ne+Ne, Si+Si, Ru+Ru, and U+U collisions. For each collision system, the upper panel shows $v_2\{2\}$, with filled and open blue squares representing the classical and quantum  results, respectively. Filled red circles denote experimental data~\cite{ATLAS:2025nnt,STAR:2021mii,STAR:2022gki}. The corresponding lower panel shows the relative difference $(Q-C)/C$, where $Q$ and $C$ denote the values of $v_2\{2\}$ obtained using quantum and classical nuclear rotors, respectively.}
\label{fig:Coverlap}
\end{figure}

In this subsection, we focus on the impact of the quantum-rotor treatment on final-state observables in Ne+Ne, Si+Si, Ru+Ru, and U+U collisions. It should be noted that all other model parameters are kept fixed when switching from the classical to the quantum rotor treatment, without additional tuning to reproduce the experimental data. The experimental data are included only as a reasonable reference, since our primary focus is on the changes from the classical to the quantum results. A more quantitative description of the experimental data may require further adjustments of the model parameters. 

Figure~\ref{fig:Coverlap} illustrates the centrality dependence of the two-particle correlation elliptic flow, $v_2\{2\}$, for the four collision systems. For each system, the $v_2\{2\}$ obtained from quantum configurations (oppen blue squares) is generally lower than or equal to that from classical configurations (filled blue squares), with the most pronounced suppression observed in central collisions, where elliptic flow is particularly sensitive to nuclear structure. This indicates that quantum superposition of nuclear orientations systematically suppresses elliptic flow.

\begin{figure}[t]
\centering
\includegraphics[width=\columnwidth]{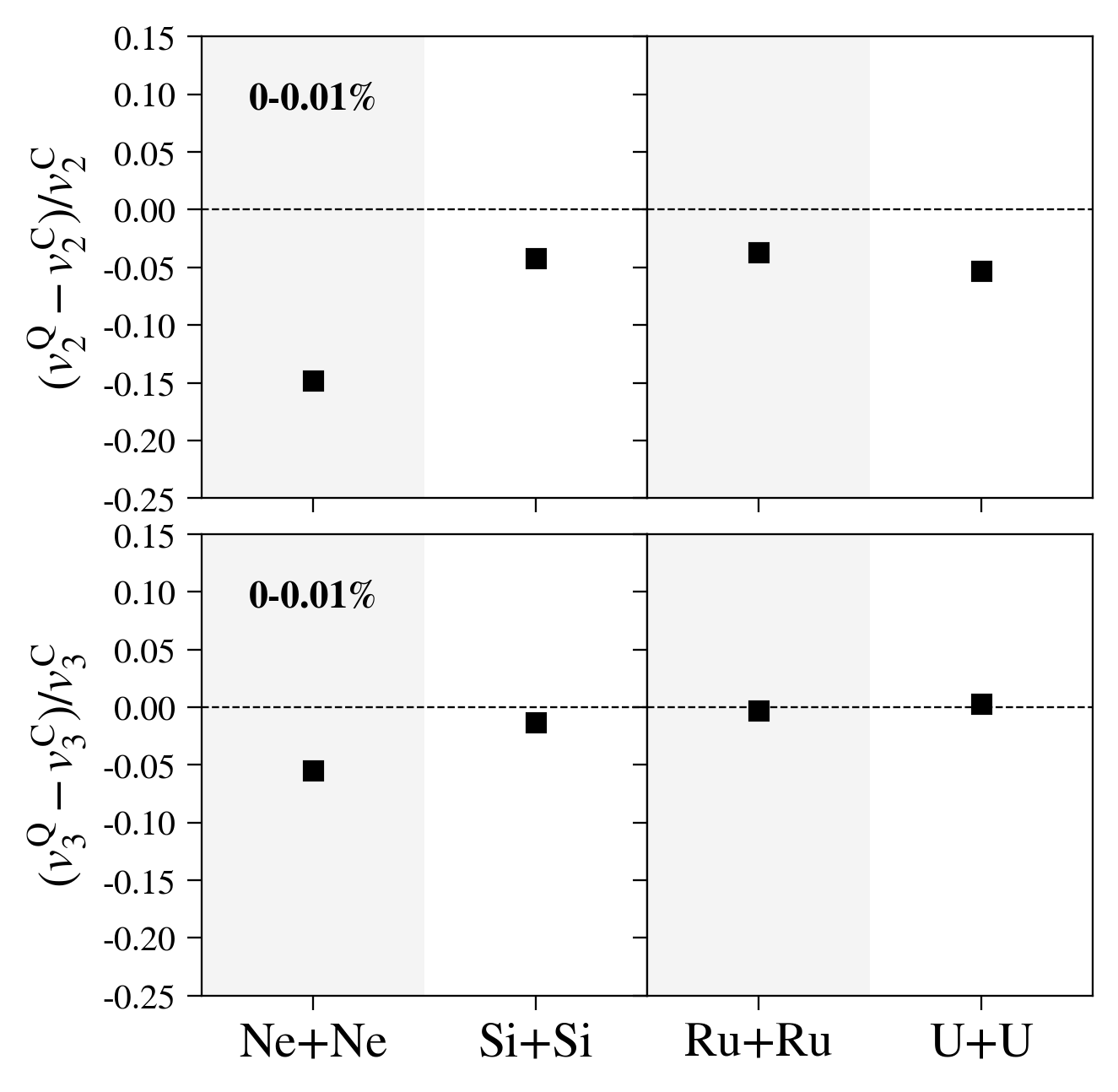}
\caption{Relative quantum corrections $(v_n\{2\}^{\mathrm{Q}}-v_n\{2\}^{\mathrm{C}})/v_n\{2\}^{\mathrm{C}}$ for $n=2$ (top) and $n=3$ (bottom) in $0$--$0.01\%$ ultra-central collisions. Ne+Ne and Si+Si are at $\sqrt{s_{NN}}=5.36$ TeV, while Ru+Ru and U+U are at 200 GeV.}
\label{fig:qcratio_both}
\end{figure}

Therefore, we focus on the $0$--$0.01\%$ ultra-central collisions. Figure~\ref{fig:qcratio_both} displays the relative difference $(v_n\{2\}^{\mathrm{Q}}-v_n\{2\}^{\mathrm{C}})/v_n\{2\}^{\mathrm{C}}$ for $n=2, 3$. The quantum-induced suppression of $v_2\{2\}$ is most pronounced in Ne+Ne collisions, reaching approximately $14.9\%$, consistent with the strong prolate deformation of $^{20}\mathrm{Ne}$. In contrast, the suppression in Si+Si collisions is only about $4.2\%$. This trend is consistent with the expectation from the initial-state study~\cite{Ke:2025tyv}. The quantum effect remains appreciable in heavy-nucleus collisions, with $v_2\{2\}$ suppressed by approximately $3.7\%$ in Ru+Ru and $5.3\%$ in U+U. It is noteworthy that, despite the sizable quadrupole deformation of $^{28}\mathrm{Si}$ ($|\beta_2|=0.313$), the quantum-induced suppression of $v_2$ is substantially weaker in Si+Si than in Ne+Ne collisions, consistent with the opposite signs of $\beta_2$ discussed in the previous section.

\begin{figure*}[t]
\centering
\includegraphics[width=\textwidth]{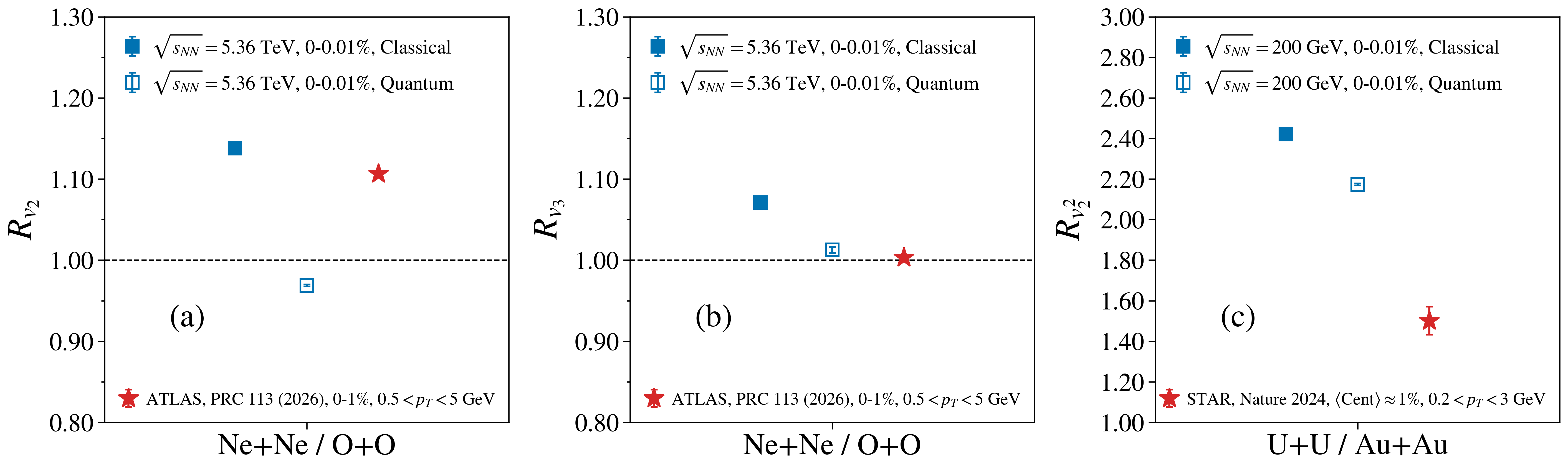}
\caption{Cross-system flow ratios in $0$--$0.01\%$ ultra-central collisions: (a) $R_{v_2}$ for Ne+Ne/O+O, (b) $R_{v_3}$ for Ne+Ne/O+O, and (c) $R_{v_2^2}$ for U+U/Au+Au. Filled and open blue squares represent the classical and quantum results, respectively, while red stars denote the corresponding experimental data~\cite{ATLAS:2025nnt,STAR:2024wgy}.}
\label{fig:NeNe_compare}
\end{figure*}

In contrast to the sizable suppression of elliptic flow, $v_3\{2\}$ exhibits only a modest suppression of approximately $5.5\%$ in Ne+Ne collisions, where $^{20}\mathrm{Ne}$ has a nonzero octupole deformation ($\beta_3=0.216$). For Si+Si, Ru+Ru, and U+U collisions, whose nuclei have no intrinsic octupole deformation in their intrinsic frames, the relative quantum corrections remain close to zero. Since triangular flow is predominantly driven by initial-state fluctuations, these results suggest that quantum superposition has only a limited impact on the fluctuation-driven component of $v_3\{2\}$.

Ratios of flow observables between collision systems of similar size can partially cancel common hydrodynamic response effects, providing a more direct probe of differences in nuclear structure. We therefore calculate the cross-system ratios
\begin{equation}
R_{v_2}
=
\frac{v_2\{2\}^{\mathrm{Ne+Ne}}}
{v_2\{2\}^{\mathrm{O+O}}},
\label{eq:Rv2_NeO}
\end{equation}

\begin{equation}
R_{v_3}
=
\frac{v_3\{2\}^{\mathrm{Ne+Ne}}}
{v_3\{2\}^{\mathrm{O+O}}},
\label{eq:Rv3_NeO}
\end{equation}

\begin{equation}
R_{v_2^2}
=
\frac{v_2^2\{2\}^{\mathrm{U+U}}}
{v_2^2\{2\}^{\mathrm{Au+Au}}}.
\label{eq:Rv22_UAu}
\end{equation}
For both classical and quantum nuclear rotors, using O+O and Au+Au as the corresponding spherical baselines. The results are shown in Fig.~\ref{fig:NeNe_compare}, together with the corresponding experimental data~\cite{ATLAS:2025nnt,STAR:2024wgy}.

As shown in Fig.~\ref{fig:NeNe_compare}(a), quantum superposition leads to a substantial reduction of $R_{v_2}$, from $1.138$ in the classical treatment to $0.969$ in the quantum treatment, falling below unity and contrasting with the experimental value of $1.106$~\cite{ATLAS:2025nnt}. One possible explanation for this discrepancy lies in our treatment of nuclear configurations. In the present work, we use Woods--Saxon distributions to describe intrinsic nuclear densities, without explicitly incorporating nucleon correlations associated with possible $\alpha$-cluster structures in $^{16}\mathrm{O}$ and $^{20}\mathrm{Ne}$~\cite{Mehrabpour:2025ogw,Li:2025hae,MenonKavumpadikkalRadhakrishnan:2026tfn}. Within this framework, the strong suppression of geometric anisotropy by orientation quantum superposition, as shown in Fig.~\ref{fig:eps2_geom}, may leave fluctuations as the dominant source of $v_2$. With fewer nucleons, O+O collisions can exhibit larger relative initial-state fluctuations than Ne+Ne collisions, potentially leading to a larger $v_2$. A more realistic description incorporating both $\alpha$-cluster correlations and quantum superposition may modify the flow ratio and help resolve the discrepancy with experimental measurements.

The corresponding triangular-flow ratio, $R_{v_3}$, is shown in Fig.~\ref{fig:NeNe_compare}(b). Quantum superposition reduces $R_{v_3}$ from approximately $1.07$ in the classical treatment to $1.01$, bringing it closer to the ATLAS measurement, which is near unity~\cite{ATLAS:2025nnt}. Similarly, the ratio $R_{v_2^2}$ for U+U/Au+Au collisions, shown in Fig.~\ref{fig:NeNe_compare}(c), is also reduced by quantum superposition, moving closer to the STAR measurement. These results highlight the impact of quantum superposition on cross-system flow ratios. Nevertheless, the experimental data correspond to different centrality selections ($0$--$1\%$ for ATLAS and approximately $1\%$ on average for STAR), whereas our calculations focus on the $0$--$0.01\%$ ultra-central interval.

\begin{figure}[t]
\centering
\includegraphics[width=\columnwidth]{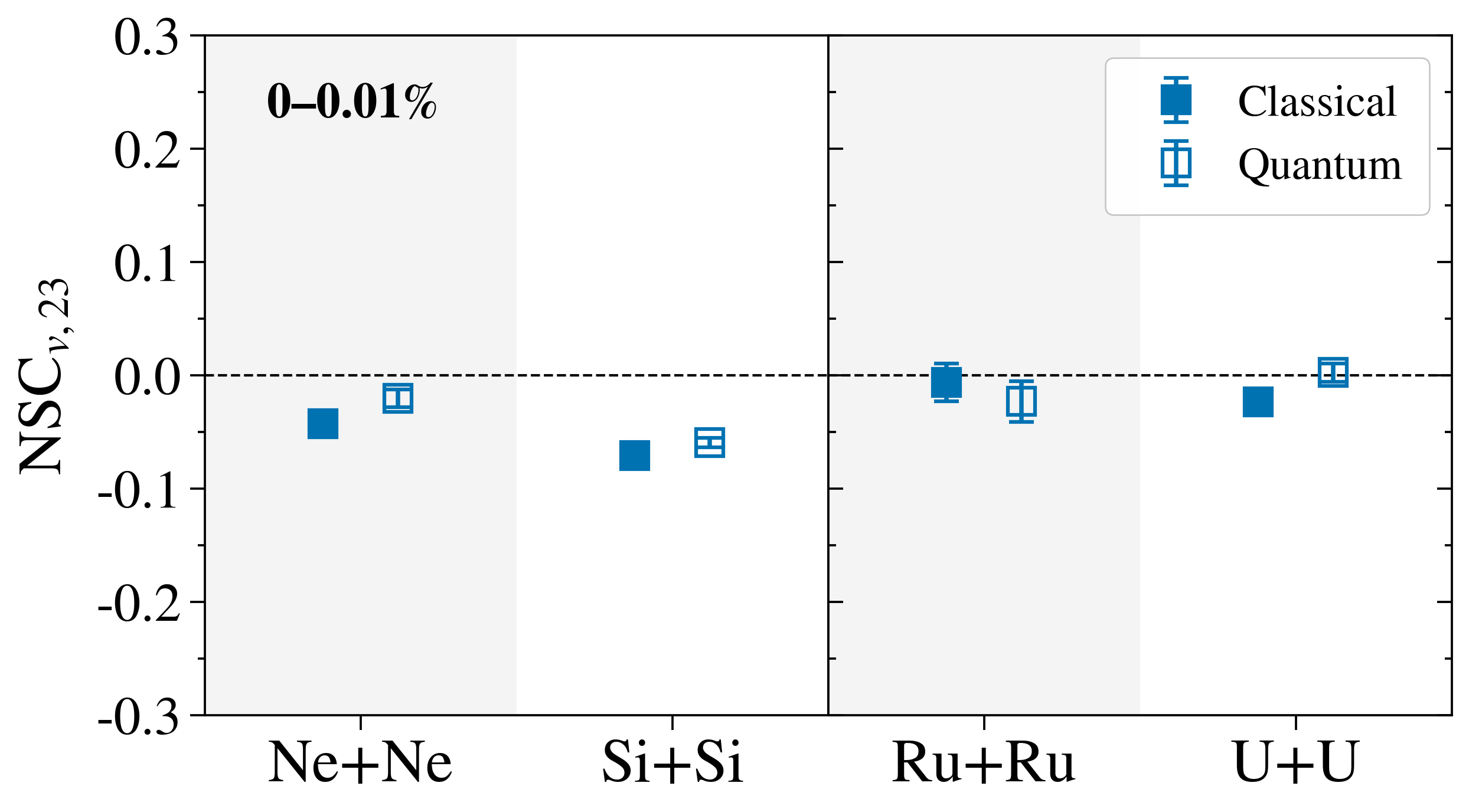}
\caption{Normalized symmetric cumulants $\mathrm{NSC}_{v,23}$ in $0$--$0.01\%$ ultra-central collisions for Ne+Ne and Si+Si at $\sqrt{s_{NN}}=5.36$ TeV, and Ru+Ru and U+U at $200$ GeV. Filled and open blue squares represent the classical and quantum results, respectively.}
\label{fig:nsc_combined}
\end{figure}

We further investigate the impact of quantum superposition on correlations between different flow harmonics. Figure~\ref{fig:nsc_combined} presents $\mathrm{NSC}_{v,23}$ for the classical and quantum configurations in four collision systems at $0$--$0.01\%$ centrality. The normalized symmetric cumulants are sensitive to nuclear structure~\cite{Li:2025hae,Wang:2024mce,Magdy:2024ooh,Magdy:2022cvt}. In the classical treatment, $\mathrm{NSC}_{v,23}$ is negative for all four systems, indicating an anticorrelation between elliptic and triangular flow. Except for Ru+Ru, where the negative correlation becomes stronger, quantum superposition weakens the anticorrelations in the other systems, bringing their $\mathrm{NSC}_{v,23}$ values closer to zero.

\begin{figure}[t]
\centering
\includegraphics[width=\columnwidth]{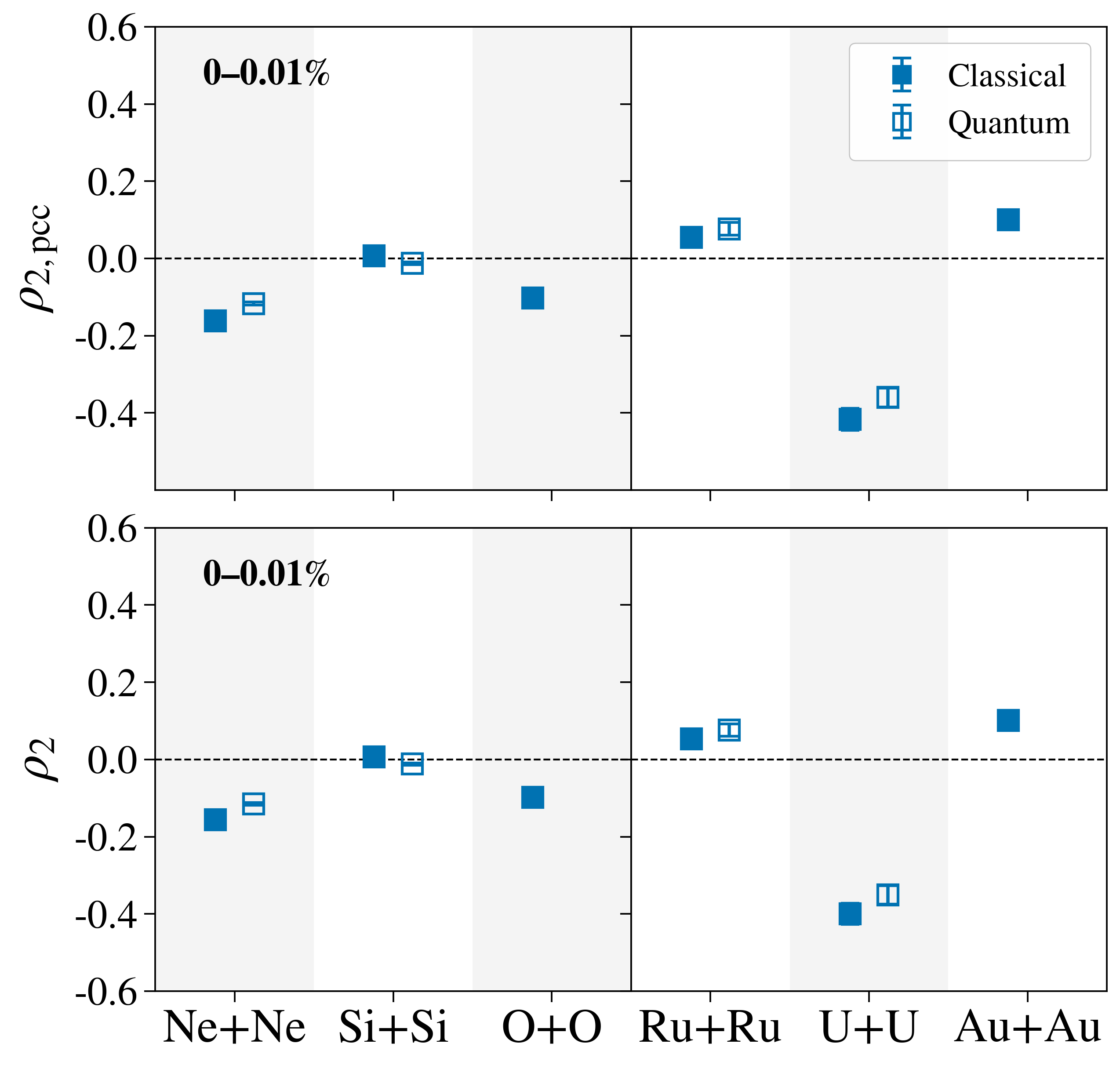}
\caption{Pearson correlation coefficients between $v_2^2$ and $[p_T]$ in $0$--$0.01\%$ ultra-central collisions. Ne+Ne, Si+Si, and O+O are at $\sqrt{s_{NN}}=5.36$ TeV, while Ru+Ru, U+U, and Au+Au are at $200$ GeV. The upper and lower panels show $\rho_{2,\mathrm{pcc}}$ and $\rho_2$. Filled and open blue squares represent the classical and quantum results, respectively.}
\label{fig:pearson_combined}
\end{figure}

\begin{figure}[t]
\centering
\includegraphics[width=\columnwidth]{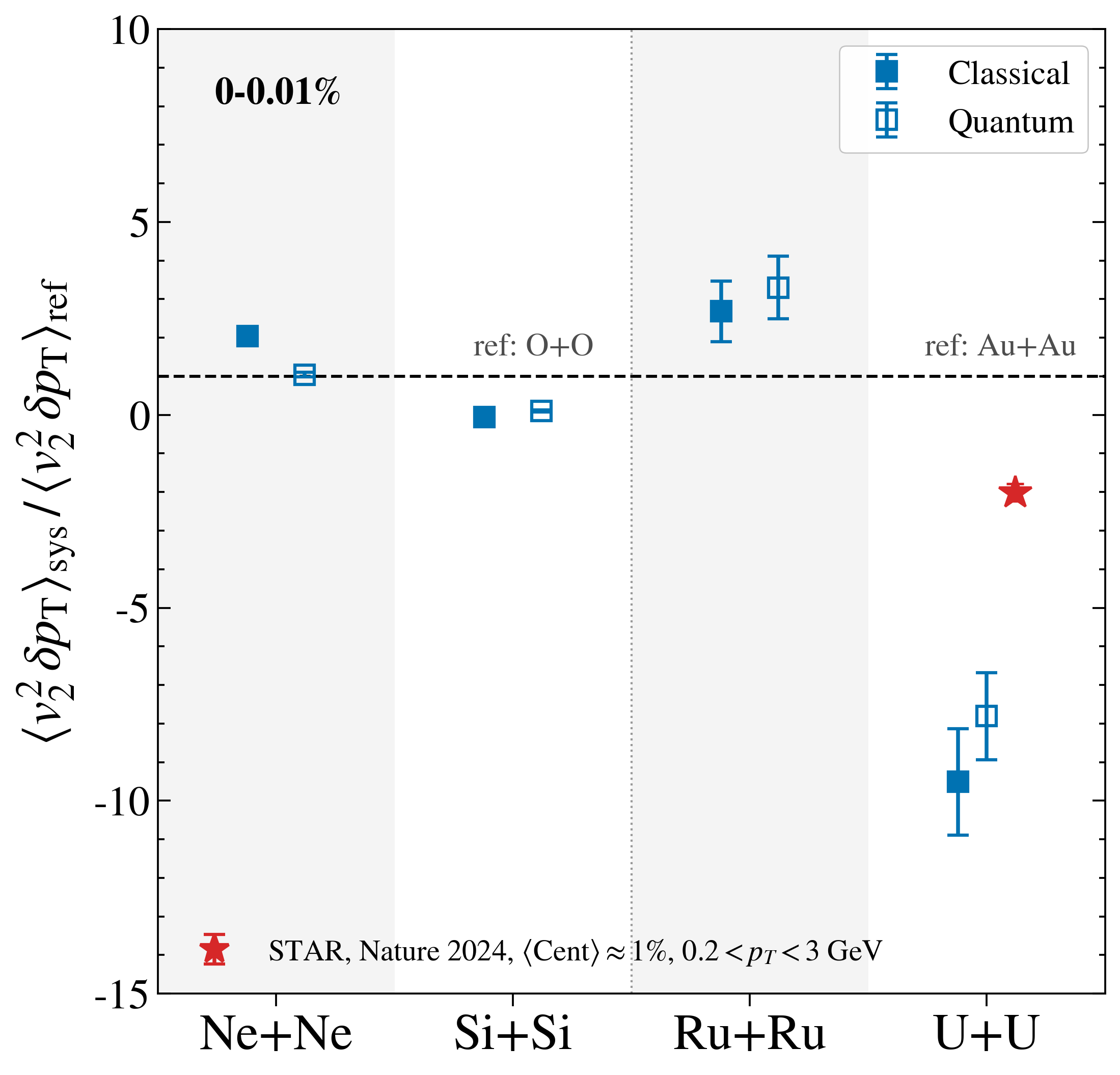}
\caption{Cross-system ratio
$R=\langle v_2^2\,\delta p_T\rangle_{\mathrm{sys}}/
\langle v_2^2\,\delta p_T\rangle_{\mathrm{ref}}$
in the $0$--$0.01\%$ most-central collisions.
O+O and Au+Au are used as the near-spherical references for the light-
and heavy-ion systems, respectively. Filled and open blue squares denote the
classical and quantum results.}
\label{fig:pearson_R}
\end{figure}

Finally, we study the correlation between $v_2^2$ and $[p_T]$ in the $0$--$0.01\%$ most-central collisions, as shown in Fig.~\ref{fig:pearson_combined}. The two definitions give very similar results in this high-multiplicity interval, consistent with suppressed finite-multiplicity effects~\cite{ATLAS:2019pvn}. They are therefore not distinguished in the following discussion, and are denoted simply by $\rho_2$. The near-spherical reference systems exhibit different baseline behaviors: O+O shows a negative correlation, whereas Au+Au gives a small positive correlation. Their difference may reflect the combined effects of system size, collision energy, and initial-state fluctuations~\cite{Giacalone:2020byk}. Here, O+O and Au+Au are used only as near-spherical references for the
corresponding light- and heavy-ion collision systems.

As shown in Fig.~\ref{fig:pearson_combined}, for the light systems, the quantum results for Ne+Ne and Si+Si are closer to the O+O reference than the corresponding classical results. A similar trend is observed for the heavy systems, where the quantum results for Ru+Ru and U+U move toward the Au+Au reference. This behavior can be qualitatively understood from the parametrization proposed in Ref.~\cite{Jia:2021wbq}. In ultra-central collisions, the Pearson coefficient approximately follows
\begin{equation}
\rho(v_2^2,[p_T])
\simeq b_1+b_2\beta_2,
\label{eq:rho_deform}
\end{equation}
where $b_1$ represents the spherical baseline and $b_2<0$ characterizes the linear dependence on the signed quadrupole deformation. Consequently, prolate deformation ($\beta_2>0$) shifts the correlation below the spherical baseline, whereas oblate deformation ($\beta_2<0$) shifts it above the baseline. The trend as shown in Fig.~\ref{fig:pearson_combined} suggests that quantum superposition weakens the effective contribution of the deformation-dependent term $b_2\beta_2$, bringing the Pearson coefficients closer to their corresponding spherical baselines.

To further illustrate the impact of quantum superposition on the deformation-sensitive covariance, Fig.~\ref{fig:pearson_R} shows the cross-system ratio
\begin{equation}
\frac{\langle v_2^2\,\delta p_T\rangle_{\mathrm{sys}}}
{\langle v_2^2\,\delta p_T\rangle_{\mathrm{ref}}},
\end{equation}
where O+O and Au+Au are used as the near-spherical references for the light- and heavy-ion systems, respectively. For the prolate systems, Ne+Ne, Ru+Ru, and U+U, quantum superposition produces noticeable but system-dependent changes in the covariance ratios. At the same time, the oblate Si+Si system exhibits only a small change. 

For the heavy-ion systems, the Au+Au reference covariance obtained with an undeformed $^{197}\rm{Au}$ nucleus is close to zero, making the corresponding ratios particularly sensitive to statistical uncertainties. We expect that incorporating a realistic deformation of Au could modify the reference covariance and potentially reduce this sensitivity. In particular, the U+U/Au+Au covariance ratio has been used experimentally to constrain nuclear deformation parameters. However, our model results show that quantum superposition shifts this ratio from approximately $-9.5$ to $-7.8$. Although our calculations and the STAR measurement correspond to different centrality selections, the observed quantum correction suggests that quantum superposition may influence the extraction of these parameters.

\section{Summary and conclusions}

In this work, we employ an extended MC Glauber model to describe both classical (fixed-orientation) and quantum (orientation-superposed) rotors, with the rotational symmetry of even-even nuclei restored in our approximate quantum treatment. By comparing the classical and quantum results, we systematically investigate the effects of quantum superposition on initial-state geometric anisotropies and final-state collective observables in high-energy nuclear collisions. Experimental data are also included in our figures as a reference. However, our primary focus is not on reproducing the experimental observables, but rather on assessing the magnitude of the corrections arising from the quantum-rotor treatment. A complete description of the experimental data would require a global fit involving adjustments of many model parameters.

In the initial state, we find the quantum correction to the mean-squared eccentricity exhibits a nonmonotonic dependence on nuclear mass number, remaining sizable for light nuclei and non-negligible even for heavy nuclei. After approximately subtracting the fluctuation contribution, the magnitude of the geometry-driven quantum correction decreases largely monotonically with increasing mass number. Furthermore, we find that quantum superposition suppresses the geometric eccentricity more strongly for prolate nuclei ($\beta_2>0$) than for oblate nuclei ($\beta_2<0$). We attribute this difference to the distinct structures of their quantum rotors, suggesting that the quantum effect depends not only on the magnitude but also on the sign of the intrinsic quadrupole deformation.

In the final state, we investigate both light- and heavy-ion collisions, including Ne+Ne and Si+Si collisions at $\sqrt{s_{NN}}=5.36$ TeV, and Ru+Ru and U+U collisions at $200$ GeV. In the $0$--$0.01\%$ centrality interval, the results show that quantum effects are most pronounced, with the strongest suppression of $v_2\{2\}$ and $v_3\{2\}$ occurring in Ne+Ne collisions, where $^{20}\mathrm{Ne}$ has the strongest prolate deformation and a nonzero octupole deformation. In contrast, the influence on $v_3\{2\}$ is small for nuclei without intrinsic octupole deformation, suggesting a limited impact on the fluctuation-driven component of triangular flow. Quantum superposition also modifies correlations between different flow harmonics, leading to system-dependent changes in $\mathrm{NSC}_{v,23}$. For the Pearson correlation between $v_2^2$ and $[p_T]$, a common trend emerges in both light- and heavy-ion systems: the quantum results move closer to their corresponding near-spherical reference values, suggesting a reduced contribution from nuclear deformation.

Furthermore, quantum superposition affects cross-system flow ratios, including the U+U/Au+Au covariance ratio used experimentally to constrain nuclear deformation parameters. Our results suggest that quantum nuclear configurations may therefore influence the extraction of these parameters from heavy-ion collision measurements. Nevertheless, the present framework does not explicitly incorporate nucleon correlations associated with possible $\alpha$-cluster structures in $^{16}\mathrm{O}$ and $^{20}\mathrm{Ne}$, and treats $^{16}\mathrm{O}$ and $^{197}\mathrm{Au}$ as undeformed reference nuclei, although their realistic nuclear structures may deviate from this spherical approximation. Future studies incorporating more realistic nuclear structures, including cluster correlations and nuclear deformations, will help clarify how quantum superposition and nuclear structure jointly influence collective observables.

\begin{acknowledgments}
The author would like to thank Giuliano Giacalone, Jiangyong Jia, Huichao Song, Qun Wang, Hao-jie Xu, Shuai Yang, Jiangming Yao, Chunjian Zhang, Pengwei Zhao, Weibin Zhao, Bo Zhou, and for discussions and comments on the manuscript. Special thanks to Jiangyong Jia, Huichao Song, Chunjian Zhang for their kind invitation to a series of workshops that brought together the nuclear structure and relativistic heavy-ion collision communities, fostering the development of this work. The authors acknowledge the use of an AI-assisted language tool to improve the clarity and grammar of the manuscript. This work is supported by NSFC under Grant No. 12575140.
\end{acknowledgments}

\bibliography{references}

\begin{thebibliography}{63}%
\makeatletter
\providecommand \@ifxundefined [1]{%
 \@ifx{#1\undefined}
}%
\providecommand \@ifnum [1]{%
 \ifnum #1\expandafter \@firstoftwo
 \else \expandafter \@secondoftwo
 \fi
}%
\providecommand \@ifx [1]{%
 \ifx #1\expandafter \@firstoftwo
 \else \expandafter \@secondoftwo
 \fi
}%
\providecommand \natexlab [1]{#1}%
\providecommand \enquote  [1]{``#1''}%
\providecommand \bibnamefont  [1]{#1}%
\providecommand \bibfnamefont [1]{#1}%
\providecommand \citenamefont [1]{#1}%
\providecommand \href@noop [0]{\@secondoftwo}%
\providecommand \href [0]{\begingroup \@sanitize@url \@href}%
\providecommand \@href[1]{\@@startlink{#1}\@@href}%
\providecommand \@@href[1]{\endgroup#1\@@endlink}%
\providecommand \@sanitize@url [0]{\catcode `\\12\catcode `\$12\catcode `\&12\catcode `\#12\catcode `\^12\catcode `\_12\catcode `\%12\relax}%
\providecommand \@@startlink[1]{}%
\providecommand \@@endlink[0]{}%
\providecommand \url  [0]{\begingroup\@sanitize@url \@url }%
\providecommand \@url [1]{\endgroup\@href {#1}{\urlprefix }}%
\providecommand \urlprefix  [0]{URL }%
\providecommand \Eprint [0]{\href }%
\providecommand \doibase [0]{https://doi.org/}%
\providecommand \selectlanguage [0]{\@gobble}%
\providecommand \bibinfo  [0]{\@secondoftwo}%
\providecommand \bibfield  [0]{\@secondoftwo}%
\providecommand \translation [1]{[#1]}%
\providecommand \BibitemOpen [0]{}%
\providecommand \bibitemStop [0]{}%
\providecommand \bibitemNoStop [0]{.\EOS\space}%
\providecommand \EOS [0]{\spacefactor3000\relax}%
\providecommand \BibitemShut  [1]{\csname bibitem#1\endcsname}%
\let\auto@bib@innerbib\@empty
\bibitem [{\citenamefont {Busza}\ \emph {et~al.}(2018)\citenamefont {Busza}, \citenamefont {Rajagopal},\ and\ \citenamefont {van~der Schee}}]{Busza:2018rrf}%
  \BibitemOpen
  \bibfield  {author} {\bibinfo {author} {\bibfnamefont {W.}~\bibnamefont {Busza}}, \bibinfo {author} {\bibfnamefont {K.}~\bibnamefont {Rajagopal}},\ and\ \bibinfo {author} {\bibfnamefont {W.}~\bibnamefont {van~der Schee}},\ }\bibfield  {title} {\bibinfo {title} {{Heavy Ion Collisions: The Big Picture, and the Big Questions}},\ }\href {https://doi.org/10.1146/annurev-nucl-101917-020852} {\bibfield  {journal} {\bibinfo  {journal} {Ann. Rev. Nucl. Part. Sci.}\ }\textbf {\bibinfo {volume} {68}},\ \bibinfo {pages} {339} (\bibinfo {year} {2018})},\ \Eprint {https://arxiv.org/abs/1802.04801} {arXiv:1802.04801 [hep-ph]} \BibitemShut {NoStop}%
\bibitem [{\citenamefont {He}\ \emph {et~al.}(2023)\citenamefont {He}, \citenamefont {van Hees},\ and\ \citenamefont {Rapp}}]{He:2022ywp}%
  \BibitemOpen
  \bibfield  {author} {\bibinfo {author} {\bibfnamefont {M.}~\bibnamefont {He}}, \bibinfo {author} {\bibfnamefont {H.}~\bibnamefont {van Hees}},\ and\ \bibinfo {author} {\bibfnamefont {R.}~\bibnamefont {Rapp}},\ }\bibfield  {title} {\bibinfo {title} {{Heavy-quark diffusion in the quark{\textendash}gluon plasma}},\ }\href {https://doi.org/10.1016/j.ppnp.2023.104020} {\bibfield  {journal} {\bibinfo  {journal} {Prog. Part. Nucl. Phys.}\ }\textbf {\bibinfo {volume} {130}},\ \bibinfo {pages} {104020} (\bibinfo {year} {2023})},\ \Eprint {https://arxiv.org/abs/2204.09299} {arXiv:2204.09299 [hep-ph]} \BibitemShut {NoStop}%
\bibitem [{\citenamefont {Adler}\ \emph {et~al.}(2004)\citenamefont {Adler} \emph {et~al.}}]{PHENIX:2003iij}%
  \BibitemOpen
  \bibfield  {author} {\bibinfo {author} {\bibfnamefont {S.~S.}\ \bibnamefont {Adler}} \emph {et~al.} (\bibinfo {collaboration} {PHENIX}),\ }\bibfield  {title} {\bibinfo {title} {{Identified charged particle spectra and yields in {Au+Au} collisions at {$\sqrt{s_{NN}}=200\,\mathrm{GeV}$}}},\ }\href {https://doi.org/10.1103/PhysRevC.69.034909} {\bibfield  {journal} {\bibinfo  {journal} {Phys. Rev. C}\ }\textbf {\bibinfo {volume} {69}},\ \bibinfo {pages} {034909} (\bibinfo {year} {2004})},\ \Eprint {https://arxiv.org/abs/nucl-ex/0307022} {arXiv:nucl-ex/0307022} \BibitemShut {NoStop}%
\bibitem [{\citenamefont {Gale}\ \emph {et~al.}(2013)\citenamefont {Gale}, \citenamefont {Jeon}, \citenamefont {Schenke}, \citenamefont {Tribedy},\ and\ \citenamefont {Venugopalan}}]{Gale:2012rq}%
  \BibitemOpen
  \bibfield  {author} {\bibinfo {author} {\bibfnamefont {C.}~\bibnamefont {Gale}}, \bibinfo {author} {\bibfnamefont {S.}~\bibnamefont {Jeon}}, \bibinfo {author} {\bibfnamefont {B.}~\bibnamefont {Schenke}}, \bibinfo {author} {\bibfnamefont {P.}~\bibnamefont {Tribedy}},\ and\ \bibinfo {author} {\bibfnamefont {R.}~\bibnamefont {Venugopalan}},\ }\bibfield  {title} {\bibinfo {title} {{Event-by-event anisotropic flow in heavy-ion collisions from combined Yang-Mills and viscous fluid dynamics}},\ }\href {https://doi.org/10.1103/PhysRevLett.110.012302} {\bibfield  {journal} {\bibinfo  {journal} {Phys. Rev. Lett.}\ }\textbf {\bibinfo {volume} {110}},\ \bibinfo {pages} {012302} (\bibinfo {year} {2013})},\ \Eprint {https://arxiv.org/abs/1209.6330} {arXiv:1209.6330 [nucl-th]} \BibitemShut {NoStop}%
\bibitem [{\citenamefont {Zhao}\ \emph {et~al.}(2017)\citenamefont {Zhao}, \citenamefont {Xu},\ and\ \citenamefont {Song}}]{Zhao:2017yhj}%
  \BibitemOpen
  \bibfield  {author} {\bibinfo {author} {\bibfnamefont {W.}~\bibnamefont {Zhao}}, \bibinfo {author} {\bibfnamefont {H.-j.}\ \bibnamefont {Xu}},\ and\ \bibinfo {author} {\bibfnamefont {H.}~\bibnamefont {Song}},\ }\bibfield  {title} {\bibinfo {title} {{Collective flow in 2.76 TeV and 5.02 A TeV Pb+Pb collisions}},\ }\href {https://doi.org/10.1140/epjc/s10052-017-5186-x} {\bibfield  {journal} {\bibinfo  {journal} {Eur. Phys. J. C}\ }\textbf {\bibinfo {volume} {77}},\ \bibinfo {pages} {645} (\bibinfo {year} {2017})},\ \Eprint {https://arxiv.org/abs/1703.10792} {arXiv:1703.10792 [nucl-th]} \BibitemShut {NoStop}%
\bibitem [{\citenamefont {Nielsen}\ and\ \citenamefont {Zhou}(2023)}]{Nielsen:2022jms}%
  \BibitemOpen
  \bibfield  {author} {\bibinfo {author} {\bibfnamefont {E.~G.}\ \bibnamefont {Nielsen}}\ and\ \bibinfo {author} {\bibfnamefont {Y.}~\bibnamefont {Zhou}},\ }\bibfield  {title} {\bibinfo {title} {{Transverse momentum decorrelation of the flow vector in Pb{\textendash}Pb collisions at $\sqrt{s_{_{\textrm{NN}}}}$ = 5.02 TeV}},\ }\href {https://doi.org/10.1140/epjc/s10052-023-11693-7} {\bibfield  {journal} {\bibinfo  {journal} {Eur. Phys. J. C}\ }\textbf {\bibinfo {volume} {83}},\ \bibinfo {pages} {545} (\bibinfo {year} {2023})},\ \Eprint {https://arxiv.org/abs/2211.13651} {arXiv:2211.13651 [nucl-ex]} \BibitemShut {NoStop}%
\bibitem [{\citenamefont {Heinz}\ and\ \citenamefont {Snellings}(2013)}]{Heinz:2013th}%
  \BibitemOpen
  \bibfield  {author} {\bibinfo {author} {\bibfnamefont {U.}~\bibnamefont {Heinz}}\ and\ \bibinfo {author} {\bibfnamefont {R.}~\bibnamefont {Snellings}},\ }\bibfield  {title} {\bibinfo {title} {{Collective flow and viscosity in relativistic heavy-ion collisions}},\ }\href {https://doi.org/10.1146/annurev-nucl-102212-170540} {\bibfield  {journal} {\bibinfo  {journal} {Ann. Rev. Nucl. Part. Sci.}\ }\textbf {\bibinfo {volume} {63}},\ \bibinfo {pages} {123} (\bibinfo {year} {2013})},\ \Eprint {https://arxiv.org/abs/1301.2826} {arXiv:1301.2826 [nucl-th]} \BibitemShut {NoStop}%
\bibitem [{\citenamefont {Everett}\ \emph {et~al.}(2021{\natexlab{a}})\citenamefont {Everett} \emph {et~al.}}]{JETSCAPE:2020shq}%
  \BibitemOpen
  \bibfield  {author} {\bibinfo {author} {\bibfnamefont {D.}~\bibnamefont {Everett}} \emph {et~al.} (\bibinfo {collaboration} {JETSCAPE}),\ }\bibfield  {title} {\bibinfo {title} {{Phenomenological constraints on the transport properties of QCD matter with data-driven model averaging}},\ }\href {https://doi.org/10.1103/PhysRevLett.126.242301} {\bibfield  {journal} {\bibinfo  {journal} {Phys. Rev. Lett.}\ }\textbf {\bibinfo {volume} {126}},\ \bibinfo {pages} {242301} (\bibinfo {year} {2021}{\natexlab{a}})},\ \Eprint {https://arxiv.org/abs/2010.03928} {arXiv:2010.03928 [hep-ph]} \BibitemShut {NoStop}%
\bibitem [{\citenamefont {Everett}\ \emph {et~al.}(2021{\natexlab{b}})\citenamefont {Everett} \emph {et~al.}}]{JETSCAPE:2020mzn}%
  \BibitemOpen
  \bibfield  {author} {\bibinfo {author} {\bibfnamefont {D.}~\bibnamefont {Everett}} \emph {et~al.} (\bibinfo {collaboration} {JETSCAPE}),\ }\bibfield  {title} {\bibinfo {title} {{Multisystem Bayesian constraints on the transport coefficients of QCD matter}},\ }\href {https://doi.org/10.1103/PhysRevC.103.054904} {\bibfield  {journal} {\bibinfo  {journal} {Phys. Rev. C}\ }\textbf {\bibinfo {volume} {103}},\ \bibinfo {pages} {054904} (\bibinfo {year} {2021}{\natexlab{b}})},\ \Eprint {https://arxiv.org/abs/2011.01430} {arXiv:2011.01430 [hep-ph]} \BibitemShut {NoStop}%
\bibitem [{\citenamefont {Parkkila}\ \emph {et~al.}(2022)\citenamefont {Parkkila}, \citenamefont {Onnerstad}, \citenamefont {Taghavi}, \citenamefont {Mordasini}, \citenamefont {Bilandzic}, \citenamefont {Virta},\ and\ \citenamefont {Kim}}]{Parkkila:2021yha}%
  \BibitemOpen
  \bibfield  {author} {\bibinfo {author} {\bibfnamefont {J.~E.}\ \bibnamefont {Parkkila}}, \bibinfo {author} {\bibfnamefont {A.}~\bibnamefont {Onnerstad}}, \bibinfo {author} {\bibfnamefont {S.~F.}\ \bibnamefont {Taghavi}}, \bibinfo {author} {\bibfnamefont {C.}~\bibnamefont {Mordasini}}, \bibinfo {author} {\bibfnamefont {A.}~\bibnamefont {Bilandzic}}, \bibinfo {author} {\bibfnamefont {M.}~\bibnamefont {Virta}},\ and\ \bibinfo {author} {\bibfnamefont {D.~J.}\ \bibnamefont {Kim}},\ }\bibfield  {title} {\bibinfo {title} {{New constraints for QCD matter from improved Bayesian parameter estimation in heavy-ion collisions at LHC}},\ }\href {https://doi.org/10.1016/j.physletb.2022.137485} {\bibfield  {journal} {\bibinfo  {journal} {Phys. Lett. B}\ }\textbf {\bibinfo {volume} {835}},\ \bibinfo {pages} {137485} (\bibinfo {year} {2022})},\ \Eprint {https://arxiv.org/abs/2111.08145} {arXiv:2111.08145 [hep-ph]} \BibitemShut {NoStop}%
\bibitem [{\citenamefont {Giacalone}(2020{\natexlab{a}})}]{Giacalone:2019pca}%
  \BibitemOpen
  \bibfield  {author} {\bibinfo {author} {\bibfnamefont {G.}~\bibnamefont {Giacalone}},\ }\bibfield  {title} {\bibinfo {title} {{Observing the deformation of nuclei with relativistic nuclear collisions}},\ }\href {https://doi.org/10.1103/PhysRevLett.124.202301} {\bibfield  {journal} {\bibinfo  {journal} {Phys. Rev. Lett.}\ }\textbf {\bibinfo {volume} {124}},\ \bibinfo {pages} {202301} (\bibinfo {year} {2020}{\natexlab{a}})},\ \Eprint {https://arxiv.org/abs/1910.04673} {arXiv:1910.04673 [nucl-th]} \BibitemShut {NoStop}%
\bibitem [{\citenamefont {Giacalone}(2020{\natexlab{b}})}]{Giacalone:2020awm}%
  \BibitemOpen
  \bibfield  {author} {\bibinfo {author} {\bibfnamefont {G.}~\bibnamefont {Giacalone}},\ }\bibfield  {title} {\bibinfo {title} {{Constraining the quadrupole deformation of atomic nuclei with relativistic nuclear collisions}},\ }\href {https://doi.org/10.1103/PhysRevC.102.024901} {\bibfield  {journal} {\bibinfo  {journal} {Phys. Rev. C}\ }\textbf {\bibinfo {volume} {102}},\ \bibinfo {pages} {024901} (\bibinfo {year} {2020}{\natexlab{b}})},\ \Eprint {https://arxiv.org/abs/2004.14463} {arXiv:2004.14463 [nucl-th]} \BibitemShut {NoStop}%
\bibitem [{\citenamefont {Jia}(2022{\natexlab{a}})}]{Jia:2021tzt}%
  \BibitemOpen
  \bibfield  {author} {\bibinfo {author} {\bibfnamefont {J.}~\bibnamefont {Jia}},\ }\bibfield  {title} {\bibinfo {title} {{Shape of atomic nuclei in heavy ion collisions}},\ }\href {https://doi.org/10.1103/PhysRevC.105.014905} {\bibfield  {journal} {\bibinfo  {journal} {Phys. Rev. C}\ }\textbf {\bibinfo {volume} {105}},\ \bibinfo {pages} {014905} (\bibinfo {year} {2022}{\natexlab{a}})},\ \Eprint {https://arxiv.org/abs/2106.08768} {arXiv:2106.08768 [nucl-th]} \BibitemShut {NoStop}%
\bibitem [{\citenamefont {Giacalone}\ \emph {et~al.}(2021)\citenamefont {Giacalone}, \citenamefont {Jia},\ and\ \citenamefont {Zhang}}]{Giacalone:2021udy}%
  \BibitemOpen
  \bibfield  {author} {\bibinfo {author} {\bibfnamefont {G.}~\bibnamefont {Giacalone}}, \bibinfo {author} {\bibfnamefont {J.}~\bibnamefont {Jia}},\ and\ \bibinfo {author} {\bibfnamefont {C.}~\bibnamefont {Zhang}},\ }\bibfield  {title} {\bibinfo {title} {{Impact of Nuclear Deformation on Relativistic Heavy-Ion Collisions: Assessing Consistency in Nuclear Physics across Energy Scales}},\ }\href {https://doi.org/10.1103/PhysRevLett.127.242301} {\bibfield  {journal} {\bibinfo  {journal} {Phys. Rev. Lett.}\ }\textbf {\bibinfo {volume} {127}},\ \bibinfo {pages} {242301} (\bibinfo {year} {2021})},\ \Eprint {https://arxiv.org/abs/2105.01638} {arXiv:2105.01638 [nucl-th]} \BibitemShut {NoStop}%
\bibitem [{\citenamefont {Jia}(2022{\natexlab{b}})}]{Jia:2021qyu}%
  \BibitemOpen
  \bibfield  {author} {\bibinfo {author} {\bibfnamefont {J.}~\bibnamefont {Jia}},\ }\bibfield  {title} {\bibinfo {title} {{Probing triaxial deformation of atomic nuclei in high-energy heavy ion collisions}},\ }\href {https://doi.org/10.1103/PhysRevC.105.044905} {\bibfield  {journal} {\bibinfo  {journal} {Phys. Rev. C}\ }\textbf {\bibinfo {volume} {105}},\ \bibinfo {pages} {044905} (\bibinfo {year} {2022}{\natexlab{b}})},\ \Eprint {https://arxiv.org/abs/2109.00604} {arXiv:2109.00604 [nucl-th]} \BibitemShut {NoStop}%
\bibitem [{\citenamefont {Jia}\ \emph {et~al.}(2022)\citenamefont {Jia}, \citenamefont {Huang},\ and\ \citenamefont {Zhang}}]{Jia:2021wbq}%
  \BibitemOpen
  \bibfield  {author} {\bibinfo {author} {\bibfnamefont {J.}~\bibnamefont {Jia}}, \bibinfo {author} {\bibfnamefont {S.}~\bibnamefont {Huang}},\ and\ \bibinfo {author} {\bibfnamefont {C.}~\bibnamefont {Zhang}},\ }\bibfield  {title} {\bibinfo {title} {{Probing nuclear quadrupole deformation from correlation of elliptic flow and transverse momentum in heavy ion collisions}},\ }\href {https://doi.org/10.1103/PhysRevC.105.014906} {\bibfield  {journal} {\bibinfo  {journal} {Phys. Rev. C}\ }\textbf {\bibinfo {volume} {105}},\ \bibinfo {pages} {014906} (\bibinfo {year} {2022})},\ \Eprint {https://arxiv.org/abs/2105.05713} {arXiv:2105.05713 [nucl-th]} \BibitemShut {NoStop}%
\bibitem [{\citenamefont {Aad}\ \emph {et~al.}(2023)\citenamefont {Aad} \emph {et~al.}}]{ATLAS:2022dov}%
  \BibitemOpen
  \bibfield  {author} {\bibinfo {author} {\bibfnamefont {G.}~\bibnamefont {Aad}} \emph {et~al.} (\bibinfo {collaboration} {ATLAS}),\ }\bibfield  {title} {\bibinfo {title} {{Correlations between flow and transverse momentum in Xe+Xe and Pb+Pb collisions at the LHC with the ATLAS detector: A probe of the heavy-ion initial state and nuclear deformation}},\ }\href {https://doi.org/10.1103/PhysRevC.107.054910} {\bibfield  {journal} {\bibinfo  {journal} {Phys. Rev. C}\ }\textbf {\bibinfo {volume} {107}},\ \bibinfo {pages} {054910} (\bibinfo {year} {2023})},\ \Eprint {https://arxiv.org/abs/2205.00039} {arXiv:2205.00039 [nucl-ex]} \BibitemShut {NoStop}%
\bibitem [{\citenamefont {Abdulhamid}\ \emph {et~al.}(2024)\citenamefont {Abdulhamid} \emph {et~al.}}]{STAR:2024wgy}%
  \BibitemOpen
  \bibfield  {author} {\bibinfo {author} {\bibfnamefont {M.~I.}\ \bibnamefont {Abdulhamid}} \emph {et~al.} (\bibinfo {collaboration} {STAR}),\ }\bibfield  {title} {\bibinfo {title} {{Imaging shapes of atomic nuclei in high-energy nuclear collisions}},\ }\href {https://doi.org/10.1038/s41586-024-08097-2} {\bibfield  {journal} {\bibinfo  {journal} {Nature}\ }\textbf {\bibinfo {volume} {635}},\ \bibinfo {pages} {67} (\bibinfo {year} {2024})},\ \Eprint {https://arxiv.org/abs/2401.06625} {arXiv:2401.06625 [nucl-ex]} \BibitemShut {NoStop}%
\bibitem [{\citenamefont {Bernhard}(2018)}]{Bernhard:2018hnz}%
  \BibitemOpen
  \bibfield  {author} {\bibinfo {author} {\bibfnamefont {J.~E.}\ \bibnamefont {Bernhard}},\ }\emph {\bibinfo {title} {{Bayesian parameter estimation for relativistic heavy-ion collisions}}},\ \href@noop {} {Ph.D. thesis},\ \bibinfo  {school} {Duke U.} (\bibinfo {year} {2018}),\ \Eprint {https://arxiv.org/abs/1804.06469} {arXiv:1804.06469 [nucl-th]} \BibitemShut {NoStop}%
\bibitem [{\citenamefont {Shen}\ \emph {et~al.}(2016)\citenamefont {Shen}, \citenamefont {Qiu}, \citenamefont {Song}, \citenamefont {Bernhard}, \citenamefont {Bass},\ and\ \citenamefont {Heinz}}]{Shen:2014vra}%
  \BibitemOpen
  \bibfield  {author} {\bibinfo {author} {\bibfnamefont {C.}~\bibnamefont {Shen}}, \bibinfo {author} {\bibfnamefont {Z.}~\bibnamefont {Qiu}}, \bibinfo {author} {\bibfnamefont {H.}~\bibnamefont {Song}}, \bibinfo {author} {\bibfnamefont {J.}~\bibnamefont {Bernhard}}, \bibinfo {author} {\bibfnamefont {S.}~\bibnamefont {Bass}},\ and\ \bibinfo {author} {\bibfnamefont {U.}~\bibnamefont {Heinz}},\ }\bibfield  {title} {\bibinfo {title} {{The iEBE-VISHNU code package for relativistic heavy-ion collisions}},\ }\href {https://doi.org/10.1016/j.cpc.2015.08.039} {\bibfield  {journal} {\bibinfo  {journal} {Comput. Phys. Commun.}\ }\textbf {\bibinfo {volume} {199}},\ \bibinfo {pages} {61} (\bibinfo {year} {2016})},\ \Eprint {https://arxiv.org/abs/1409.8164} {arXiv:1409.8164 [nucl-th]} \BibitemShut {NoStop}%
\bibitem [{\citenamefont {Noronha-Hostler}\ \emph {et~al.}(2016)\citenamefont {Noronha-Hostler}, \citenamefont {Yan}, \citenamefont {Gardim},\ and\ \citenamefont {Ollitrault}}]{Noronha-Hostler:2015dbi}%
  \BibitemOpen
  \bibfield  {author} {\bibinfo {author} {\bibfnamefont {J.}~\bibnamefont {Noronha-Hostler}}, \bibinfo {author} {\bibfnamefont {L.}~\bibnamefont {Yan}}, \bibinfo {author} {\bibfnamefont {F.~G.}\ \bibnamefont {Gardim}},\ and\ \bibinfo {author} {\bibfnamefont {J.-Y.}\ \bibnamefont {Ollitrault}},\ }\bibfield  {title} {\bibinfo {title} {{Linear and cubic response to the initial eccentricity in heavy-ion collisions}},\ }\href {https://doi.org/10.1103/PhysRevC.93.014909} {\bibfield  {journal} {\bibinfo  {journal} {Phys. Rev. C}\ }\textbf {\bibinfo {volume} {93}},\ \bibinfo {pages} {014909} (\bibinfo {year} {2016})},\ \Eprint {https://arxiv.org/abs/1511.03896} {arXiv:1511.03896 [nucl-th]} \BibitemShut {NoStop}%
\bibitem [{\citenamefont {Abualrob}\ \emph {et~al.}(2026)\citenamefont {Abualrob} \emph {et~al.}}]{ALICE:2025luc}%
  \BibitemOpen
  \bibfield  {author} {\bibinfo {author} {\bibfnamefont {I.~J.}\ \bibnamefont {Abualrob}} \emph {et~al.} (\bibinfo {collaboration} {ALICE}),\ }\bibfield  {title} {\bibinfo {title} {{Evidence of Nuclear Geometry-Driven Anisotropic Flow in {O+O} and {Ne+Ne} Collisions at {$\sqrt{s_{NN}}=5.36\,\mathrm{TeV}$}}},\ }\href {https://doi.org/10.1103/gymp-vp87} {\bibfield  {journal} {\bibinfo  {journal} {Phys. Rev. Lett.}\ }\textbf {\bibinfo {volume} {137}},\ \bibinfo {pages} {082301} (\bibinfo {year} {2026})},\ \Eprint {https://arxiv.org/abs/2509.06428} {arXiv:2509.06428 [nucl-ex]} \BibitemShut {NoStop}%
\bibitem [{\citenamefont {{LHCb Collaboration}}(2025)}]{LHCb:2025ixz}%
  \BibitemOpen
  \bibfield  {author} {\bibinfo {author} {\bibnamefont {{LHCb Collaboration}}},\ }\href@noop {} {\emph {\bibinfo {title} {{Unveiling the shape of the $^{20}$Ne nucleus by measuring the flow coefficients with cumulants in PbNe and PbAr collisions at $\sqrt{s_{NN}} = 70.9$ GeV}}}},\ \bibinfo {type} {Tech. Rep.}\ \bibinfo {number} {LHCb-CONF-2025-001}\ (\bibinfo  {institution} {CERN},\ \bibinfo {year} {2025})\ \Eprint {https://arxiv.org/abs/2509.12399} {arXiv:2509.12399 [nucl-ex]} \BibitemShut {NoStop}%
\bibitem [{\citenamefont {Aad}\ \emph {et~al.}(2026)\citenamefont {Aad} \emph {et~al.}}]{ATLAS:2025nnt}%
  \BibitemOpen
  \bibfield  {author} {\bibinfo {author} {\bibfnamefont {G.}~\bibnamefont {Aad}} \emph {et~al.} (\bibinfo {collaboration} {ATLAS}),\ }\bibfield  {title} {\bibinfo {title} {{Measurement of the azimuthal anisotropy of charged particles in {$\sqrt{s_{NN}}=5.36\,\mathrm{TeV}$} {$^{16}\mathrm{O}+^{16}\mathrm{O}$} and {$^{20}\mathrm{Ne}+^{20}\mathrm{Ne}$} collisions with the {ATLAS} detector}},\ }\href {https://doi.org/10.1103/xqxz-8bhf} {\bibfield  {journal} {\bibinfo  {journal} {Phys. Rev. C}\ }\textbf {\bibinfo {volume} {113}},\ \bibinfo {pages} {045205} (\bibinfo {year} {2026})},\ \Eprint {https://arxiv.org/abs/2509.05171} {arXiv:2509.05171 [nucl-ex]} \BibitemShut {NoStop}%
\bibitem [{\citenamefont {Hayrapetyan}\ \emph {et~al.}(2026)\citenamefont {Hayrapetyan} \emph {et~al.}}]{CMS:2025tga}%
  \BibitemOpen
  \bibfield  {author} {\bibinfo {author} {\bibfnamefont {A.}~\bibnamefont {Hayrapetyan}} \emph {et~al.} (\bibinfo {collaboration} {CMS}),\ }\bibfield  {title} {\bibinfo {title} {{Observation of Long-Range Collective Flow in O+O and Ne+Ne Collisions and Implications for Nuclear Structure Studies}},\ }\href {https://doi.org/10.1103/26wx-tg6f} {\bibfield  {journal} {\bibinfo  {journal} {Phys. Rev. Lett.}\ }\textbf {\bibinfo {volume} {137}},\ \bibinfo {pages} {082302} (\bibinfo {year} {2026})},\ \Eprint {https://arxiv.org/abs/2510.02580} {arXiv:2510.02580 [nucl-ex]} \BibitemShut {NoStop}%
\bibitem [{\citenamefont {Giacalone}\ \emph {et~al.}(2025)\citenamefont {Giacalone} \emph {et~al.}}]{Giacalone:2024luz}%
  \BibitemOpen
  \bibfield  {author} {\bibinfo {author} {\bibfnamefont {G.}~\bibnamefont {Giacalone}} \emph {et~al.},\ }\bibfield  {title} {\bibinfo {title} {{Exploiting {$^{20}\mathrm{Ne}$} Isotopes for Precision Characterizations of Collectivity in Small Systems}},\ }\href {https://doi.org/10.1103/k8rb-jgvq} {\bibfield  {journal} {\bibinfo  {journal} {Phys. Rev. Lett.}\ }\textbf {\bibinfo {volume} {135}},\ \bibinfo {pages} {012302} (\bibinfo {year} {2025})},\ \Eprint {https://arxiv.org/abs/2402.05995} {arXiv:2402.05995 [nucl-th]} \BibitemShut {NoStop}%
\bibitem [{\citenamefont {Adam}\ \emph {et~al.}(2016{\natexlab{a}})\citenamefont {Adam} \emph {et~al.}}]{ALICE:2016ccg}%
  \BibitemOpen
  \bibfield  {author} {\bibinfo {author} {\bibfnamefont {J.}~\bibnamefont {Adam}} \emph {et~al.} (\bibinfo {collaboration} {ALICE}),\ }\bibfield  {title} {\bibinfo {title} {{Anisotropic flow of charged particles in Pb-Pb collisions at $\sqrt{s_{\rm NN}}=5.02$ TeV}},\ }\href {https://doi.org/10.1103/PhysRevLett.116.132302} {\bibfield  {journal} {\bibinfo  {journal} {Phys. Rev. Lett.}\ }\textbf {\bibinfo {volume} {116}},\ \bibinfo {pages} {132302} (\bibinfo {year} {2016}{\natexlab{a}})},\ \Eprint {https://arxiv.org/abs/1602.01119} {arXiv:1602.01119 [nucl-ex]} \BibitemShut {NoStop}%
\bibitem [{\citenamefont {Aad}\ \emph {et~al.}(2012)\citenamefont {Aad} \emph {et~al.}}]{ATLAS:2012at}%
  \BibitemOpen
  \bibfield  {author} {\bibinfo {author} {\bibfnamefont {G.}~\bibnamefont {Aad}} \emph {et~al.} (\bibinfo {collaboration} {ATLAS}),\ }\bibfield  {title} {\bibinfo {title} {{Measurement of the azimuthal anisotropy for charged particle production in $\sqrt{s_{NN}}=2.76$ TeV lead-lead collisions with the ATLAS detector}},\ }\href {https://doi.org/10.1103/PhysRevC.86.014907} {\bibfield  {journal} {\bibinfo  {journal} {Phys. Rev. C}\ }\textbf {\bibinfo {volume} {86}},\ \bibinfo {pages} {014907} (\bibinfo {year} {2012})},\ \Eprint {https://arxiv.org/abs/1203.3087} {arXiv:1203.3087 [hep-ex]} \BibitemShut {NoStop}%
\bibitem [{\citenamefont {Voloshin}\ \emph {et~al.}(2010)\citenamefont {Voloshin}, \citenamefont {Poskanzer},\ and\ \citenamefont {Snellings}}]{Voloshin:2008dg}%
  \BibitemOpen
  \bibfield  {author} {\bibinfo {author} {\bibfnamefont {S.~A.}\ \bibnamefont {Voloshin}}, \bibinfo {author} {\bibfnamefont {A.~M.}\ \bibnamefont {Poskanzer}},\ and\ \bibinfo {author} {\bibfnamefont {R.}~\bibnamefont {Snellings}},\ }\bibfield  {title} {\bibinfo {title} {{Collective phenomena in non-central nuclear collisions}},\ }\href {https://doi.org/10.1007/978-3-642-01539-7_10} {\bibfield  {journal} {\bibinfo  {journal} {Landolt-Bornstein}\ }\textbf {\bibinfo {volume} {23}},\ \bibinfo {pages} {293} (\bibinfo {year} {2010})},\ \Eprint {https://arxiv.org/abs/0809.2949} {arXiv:0809.2949 [nucl-ex]} \BibitemShut {NoStop}%
\bibitem [{\citenamefont {Shou}\ \emph {et~al.}(2015)\citenamefont {Shou}, \citenamefont {Ma}, \citenamefont {Sorensen}, \citenamefont {Tang}, \citenamefont {Videb{\ae}k},\ and\ \citenamefont {Wang}}]{Shou:2014eya}%
  \BibitemOpen
  \bibfield  {author} {\bibinfo {author} {\bibfnamefont {Q.~Y.}\ \bibnamefont {Shou}}, \bibinfo {author} {\bibfnamefont {Y.~G.}\ \bibnamefont {Ma}}, \bibinfo {author} {\bibfnamefont {P.}~\bibnamefont {Sorensen}}, \bibinfo {author} {\bibfnamefont {A.~H.}\ \bibnamefont {Tang}}, \bibinfo {author} {\bibfnamefont {F.}~\bibnamefont {Videb{\ae}k}},\ and\ \bibinfo {author} {\bibfnamefont {H.}~\bibnamefont {Wang}},\ }\bibfield  {title} {\bibinfo {title} {{Parameterization of Deformed Nuclei for Glauber Modeling in Relativistic Heavy Ion Collisions}},\ }\href {https://doi.org/10.1016/j.physletb.2015.07.078} {\bibfield  {journal} {\bibinfo  {journal} {Phys. Lett. B}\ }\textbf {\bibinfo {volume} {749}},\ \bibinfo {pages} {215} (\bibinfo {year} {2015})},\ \Eprint {https://arxiv.org/abs/1409.8375} {arXiv:1409.8375 [nucl-th]} \BibitemShut {NoStop}%
\bibitem [{\citenamefont {Miller}\ \emph {et~al.}(2007)\citenamefont {Miller}, \citenamefont {Reygers}, \citenamefont {Sanders},\ and\ \citenamefont {Steinberg}}]{Miller:2007ri}%
  \BibitemOpen
  \bibfield  {author} {\bibinfo {author} {\bibfnamefont {M.~L.}\ \bibnamefont {Miller}}, \bibinfo {author} {\bibfnamefont {K.}~\bibnamefont {Reygers}}, \bibinfo {author} {\bibfnamefont {S.~J.}\ \bibnamefont {Sanders}},\ and\ \bibinfo {author} {\bibfnamefont {P.}~\bibnamefont {Steinberg}},\ }\bibfield  {title} {\bibinfo {title} {{Glauber modeling in high energy nuclear collisions}},\ }\href {https://doi.org/10.1146/annurev.nucl.57.090506.123020} {\bibfield  {journal} {\bibinfo  {journal} {Ann. Rev. Nucl. Part. Sci.}\ }\textbf {\bibinfo {volume} {57}},\ \bibinfo {pages} {205} (\bibinfo {year} {2007})},\ \Eprint {https://arxiv.org/abs/nucl-ex/0701025} {arXiv:nucl-ex/0701025} \BibitemShut {NoStop}%
\bibitem [{\citenamefont {Moreland}\ \emph {et~al.}(2015)\citenamefont {Moreland}, \citenamefont {Bernhard},\ and\ \citenamefont {Bass}}]{Moreland:2014oya}%
  \BibitemOpen
  \bibfield  {author} {\bibinfo {author} {\bibfnamefont {J.~S.}\ \bibnamefont {Moreland}}, \bibinfo {author} {\bibfnamefont {J.~E.}\ \bibnamefont {Bernhard}},\ and\ \bibinfo {author} {\bibfnamefont {S.~A.}\ \bibnamefont {Bass}},\ }\bibfield  {title} {\bibinfo {title} {{Alternative ansatz to wounded nucleon and binary collision scaling in high-energy nuclear collisions}},\ }\href {https://doi.org/10.1103/PhysRevC.92.011901} {\bibfield  {journal} {\bibinfo  {journal} {Phys. Rev. C}\ }\textbf {\bibinfo {volume} {92}},\ \bibinfo {pages} {011901} (\bibinfo {year} {2015})},\ \Eprint {https://arxiv.org/abs/1412.4708} {arXiv:1412.4708 [nucl-th]} \BibitemShut {NoStop}%
\bibitem [{\citenamefont {Dobaczewski}\ \emph {et~al.}(2025)\citenamefont {Dobaczewski}, \citenamefont {Gade}, \citenamefont {Godbey}, \citenamefont {Janssens},\ and\ \citenamefont {Nazarewicz}}]{Dobaczewski:2025rdi}%
  \BibitemOpen
  \bibfield  {author} {\bibinfo {author} {\bibfnamefont {J.}~\bibnamefont {Dobaczewski}}, \bibinfo {author} {\bibfnamefont {A.}~\bibnamefont {Gade}}, \bibinfo {author} {\bibfnamefont {K.}~\bibnamefont {Godbey}}, \bibinfo {author} {\bibfnamefont {R.~V.~F.}\ \bibnamefont {Janssens}},\ and\ \bibinfo {author} {\bibfnamefont {W.}~\bibnamefont {Nazarewicz}},\ }\bibfield  {title} {\bibinfo {title} {{Extraction of ground-state nuclear deformations from ultrarelativistic heavy-ion collisions: Nuclear structure physics context}},\ }\href {https://doi.org/10.1103/kngg-1ccb} {\bibfield  {journal} {\bibinfo  {journal} {Phys. Rev. Res.}\ }\textbf {\bibinfo {volume} {7}},\ \bibinfo {pages} {043159} (\bibinfo {year} {2025})},\ \Eprint {https://arxiv.org/abs/2507.05208} {arXiv:2507.05208 [nucl-th]} \BibitemShut {NoStop}%
\bibitem [{\citenamefont {d'Enterria}\ and\ \citenamefont {Loizides}(2021)}]{dEnterria:2020dwq}%
  \BibitemOpen
  \bibfield  {author} {\bibinfo {author} {\bibfnamefont {D.}~\bibnamefont {d'Enterria}}\ and\ \bibinfo {author} {\bibfnamefont {C.}~\bibnamefont {Loizides}},\ }\bibfield  {title} {\bibinfo {title} {{Progress in the Glauber Model at Collider Energies}},\ }\href {https://doi.org/10.1146/annurev-nucl-102419-060007} {\bibfield  {journal} {\bibinfo  {journal} {Ann. Rev. Nucl. Part. Sci.}\ }\textbf {\bibinfo {volume} {71}},\ \bibinfo {pages} {315} (\bibinfo {year} {2021})},\ \Eprint {https://arxiv.org/abs/2011.14909} {arXiv:2011.14909 [hep-ph]} \BibitemShut {NoStop}%
\bibitem [{\citenamefont {Ke}(2025)}]{Ke:2025tyv}%
  \BibitemOpen
  \bibfield  {author} {\bibinfo {author} {\bibfnamefont {W.}~\bibnamefont {Ke}},\ }\href@noop {} {\bibinfo {title} {{From wave-function to fireball geometry: the role of a restored broken symmetry in ultra-relativistic collisions of deformed nuclei}}} (\bibinfo {year} {2025}),\ \Eprint {https://arxiv.org/abs/2509.09549} {arXiv:2509.09549 [nucl-th]} \BibitemShut {NoStop}%
\bibitem [{\citenamefont {Xu}\ and\ \citenamefont {Wang}(2026)}]{Xu:2026llz}%
  \BibitemOpen
  \bibfield  {author} {\bibinfo {author} {\bibfnamefont {H.-j.}\ \bibnamefont {Xu}}\ and\ \bibinfo {author} {\bibfnamefont {Q.}~\bibnamefont {Wang}},\ }\href@noop {} {\bibinfo {title} {{Quantum Symmetry Restoration and Emergent Effective Deformation in Relativistic Heavy-Ion Collisions}}} (\bibinfo {year} {2026}),\ \Eprint {https://arxiv.org/abs/2606.02412} {arXiv:2606.02412 [nucl-th]} \BibitemShut {NoStop}%
\bibitem [{\citenamefont {Bofos}\ \emph {et~al.}(2026)\citenamefont {Bofos}, \citenamefont {Li}, \citenamefont {Ding}, \citenamefont {Bally}, \citenamefont {Duguet}, \citenamefont {Frosini},\ and\ \citenamefont {Yao}}]{Bofos:2026nmg}%
  \BibitemOpen
  \bibfield  {author} {\bibinfo {author} {\bibfnamefont {S.}~\bibnamefont {Bofos}}, \bibinfo {author} {\bibfnamefont {Y.}~\bibnamefont {Li}}, \bibinfo {author} {\bibfnamefont {C.}~\bibnamefont {Ding}}, \bibinfo {author} {\bibfnamefont {B.}~\bibnamefont {Bally}}, \bibinfo {author} {\bibfnamefont {T.}~\bibnamefont {Duguet}}, \bibinfo {author} {\bibfnamefont {M.}~\bibnamefont {Frosini}},\ and\ \bibinfo {author} {\bibfnamefont {J.}~\bibnamefont {Yao}},\ }\href@noop {} {\bibinfo {title} {{Quantum effects in the quadrupole rotor picture of ultra-relativistic ion-ion collisions}}} (\bibinfo {year} {2026}),\ \Eprint {https://arxiv.org/abs/2605.28813} {arXiv:2605.28813 [nucl-th]} \BibitemShut {NoStop}%
\bibitem [{\citenamefont {Voloshin}\ and\ \citenamefont {Zhang}(1996)}]{Voloshin:1994mz}%
  \BibitemOpen
  \bibfield  {author} {\bibinfo {author} {\bibfnamefont {S.}~\bibnamefont {Voloshin}}\ and\ \bibinfo {author} {\bibfnamefont {Y.}~\bibnamefont {Zhang}},\ }\bibfield  {title} {\bibinfo {title} {{Flow study in relativistic nuclear collisions by Fourier expansion of Azimuthal particle distributions}},\ }\href {https://doi.org/10.1007/s002880050141} {\bibfield  {journal} {\bibinfo  {journal} {Z. Phys. C}\ }\textbf {\bibinfo {volume} {70}},\ \bibinfo {pages} {665} (\bibinfo {year} {1996})},\ \Eprint {https://arxiv.org/abs/hep-ph/9407282} {arXiv:hep-ph/9407282} \BibitemShut {NoStop}%
\bibitem [{\citenamefont {Adam}\ \emph {et~al.}(2016{\natexlab{b}})\citenamefont {Adam} \emph {et~al.}}]{ALICE:2016kpq}%
  \BibitemOpen
  \bibfield  {author} {\bibinfo {author} {\bibfnamefont {J.}~\bibnamefont {Adam}} \emph {et~al.} (\bibinfo {collaboration} {ALICE}),\ }\bibfield  {title} {\bibinfo {title} {{Correlated event-by-event fluctuations of flow harmonics in Pb-Pb collisions at $\sqrt{s_{_{\rm NN}}}=2.76$ TeV}},\ }\href {https://doi.org/10.1103/PhysRevLett.117.182301} {\bibfield  {journal} {\bibinfo  {journal} {Phys. Rev. Lett.}\ }\textbf {\bibinfo {volume} {117}},\ \bibinfo {pages} {182301} (\bibinfo {year} {2016}{\natexlab{b}})},\ \Eprint {https://arxiv.org/abs/1604.07663} {arXiv:1604.07663 [nucl-ex]} \BibitemShut {NoStop}%
\bibitem [{\citenamefont {Gardim}\ \emph {et~al.}(2017)\citenamefont {Gardim}, \citenamefont {Grassi}, \citenamefont {Luzum},\ and\ \citenamefont {Noronha-Hostler}}]{Gardim:2016nrr}%
  \BibitemOpen
  \bibfield  {author} {\bibinfo {author} {\bibfnamefont {F.~G.}\ \bibnamefont {Gardim}}, \bibinfo {author} {\bibfnamefont {F.}~\bibnamefont {Grassi}}, \bibinfo {author} {\bibfnamefont {M.}~\bibnamefont {Luzum}},\ and\ \bibinfo {author} {\bibfnamefont {J.}~\bibnamefont {Noronha-Hostler}},\ }\bibfield  {title} {\bibinfo {title} {{Hydrodynamic Predictions for Mixed Harmonic Correlations in 200 GeV Au+Au Collisions}},\ }\href {https://doi.org/10.1103/PhysRevC.95.034901} {\bibfield  {journal} {\bibinfo  {journal} {Phys. Rev. C}\ }\textbf {\bibinfo {volume} {95}},\ \bibinfo {pages} {034901} (\bibinfo {year} {2017})},\ \Eprint {https://arxiv.org/abs/1608.02982} {arXiv:1608.02982 [nucl-th]} \BibitemShut {NoStop}%
\bibitem [{\citenamefont {Zhu}\ \emph {et~al.}(2017)\citenamefont {Zhu}, \citenamefont {Zhou}, \citenamefont {Xu},\ and\ \citenamefont {Song}}]{Zhu:2016puf}%
  \BibitemOpen
  \bibfield  {author} {\bibinfo {author} {\bibfnamefont {X.}~\bibnamefont {Zhu}}, \bibinfo {author} {\bibfnamefont {Y.}~\bibnamefont {Zhou}}, \bibinfo {author} {\bibfnamefont {H.}~\bibnamefont {Xu}},\ and\ \bibinfo {author} {\bibfnamefont {H.}~\bibnamefont {Song}},\ }\bibfield  {title} {\bibinfo {title} {{Correlations of flow harmonics in 2.76A TeV Pb--Pb collisions}},\ }\href {https://doi.org/10.1103/PhysRevC.95.044902} {\bibfield  {journal} {\bibinfo  {journal} {Phys. Rev. C}\ }\textbf {\bibinfo {volume} {95}},\ \bibinfo {pages} {044902} (\bibinfo {year} {2017})},\ \Eprint {https://arxiv.org/abs/1608.05305} {arXiv:1608.05305 [nucl-th]} \BibitemShut {NoStop}%
\bibitem [{\citenamefont {Bozek}(2016)}]{Bozek:2016yoj}%
  \BibitemOpen
  \bibfield  {author} {\bibinfo {author} {\bibfnamefont {P.}~\bibnamefont {Bozek}},\ }\bibfield  {title} {\bibinfo {title} {{Transverse-momentum{\textendash}flow correlations in relativistic heavy-ion collisions}},\ }\href {https://doi.org/10.1103/PhysRevC.93.044908} {\bibfield  {journal} {\bibinfo  {journal} {Phys. Rev. C}\ }\textbf {\bibinfo {volume} {93}},\ \bibinfo {pages} {044908} (\bibinfo {year} {2016})},\ \Eprint {https://arxiv.org/abs/1601.04513} {arXiv:1601.04513 [nucl-th]} \BibitemShut {NoStop}%
\bibitem [{\citenamefont {Schenke}\ \emph {et~al.}(2020)\citenamefont {Schenke}, \citenamefont {Shen},\ and\ \citenamefont {Teaney}}]{Schenke:2020uqq}%
  \BibitemOpen
  \bibfield  {author} {\bibinfo {author} {\bibfnamefont {B.}~\bibnamefont {Schenke}}, \bibinfo {author} {\bibfnamefont {C.}~\bibnamefont {Shen}},\ and\ \bibinfo {author} {\bibfnamefont {D.}~\bibnamefont {Teaney}},\ }\bibfield  {title} {\bibinfo {title} {{Transverse momentum fluctuations and their correlation with elliptic flow in nuclear collision}},\ }\href {https://doi.org/10.1103/PhysRevC.102.034905} {\bibfield  {journal} {\bibinfo  {journal} {Phys. Rev. C}\ }\textbf {\bibinfo {volume} {102}},\ \bibinfo {pages} {034905} (\bibinfo {year} {2020})},\ \Eprint {https://arxiv.org/abs/2004.00690} {arXiv:2004.00690 [nucl-th]} \BibitemShut {NoStop}%
\bibitem [{\citenamefont {Liu}\ \emph {et~al.}(2015)\citenamefont {Liu}, \citenamefont {Shen},\ and\ \citenamefont {Heinz}}]{Liu:2015nwa}%
  \BibitemOpen
  \bibfield  {author} {\bibinfo {author} {\bibfnamefont {J.}~\bibnamefont {Liu}}, \bibinfo {author} {\bibfnamefont {C.}~\bibnamefont {Shen}},\ and\ \bibinfo {author} {\bibfnamefont {U.}~\bibnamefont {Heinz}},\ }\bibfield  {title} {\bibinfo {title} {{Pre-equilibrium evolution effects on heavy-ion collision observables}},\ }\href {https://doi.org/10.1103/PhysRevC.91.064906} {\bibfield  {journal} {\bibinfo  {journal} {Phys. Rev. C}\ }\textbf {\bibinfo {volume} {91}},\ \bibinfo {pages} {064906} (\bibinfo {year} {2015})},\ \bibinfo {note} {[Erratum: Phys.Rev.C 92, 049904 (2015)]},\ \Eprint {https://arxiv.org/abs/1504.02160} {arXiv:1504.02160 [nucl-th]} \BibitemShut {NoStop}%
\bibitem [{\citenamefont {Song}\ and\ \citenamefont {Heinz}(2008)}]{Song:2007ux}%
  \BibitemOpen
  \bibfield  {author} {\bibinfo {author} {\bibfnamefont {H.}~\bibnamefont {Song}}\ and\ \bibinfo {author} {\bibfnamefont {U.~W.}\ \bibnamefont {Heinz}},\ }\bibfield  {title} {\bibinfo {title} {{Causal viscous hydrodynamics in 2+1 dimensions for relativistic heavy-ion collisions}},\ }\href {https://doi.org/10.1103/PhysRevC.77.064901} {\bibfield  {journal} {\bibinfo  {journal} {Phys. Rev. C}\ }\textbf {\bibinfo {volume} {77}},\ \bibinfo {pages} {064901} (\bibinfo {year} {2008})},\ \Eprint {https://arxiv.org/abs/0712.3715} {arXiv:0712.3715 [nucl-th]} \BibitemShut {NoStop}%
\bibitem [{\citenamefont {Cooper}\ and\ \citenamefont {Frye}(1974)}]{Cooper:1974mv}%
  \BibitemOpen
  \bibfield  {author} {\bibinfo {author} {\bibfnamefont {F.}~\bibnamefont {Cooper}}\ and\ \bibinfo {author} {\bibfnamefont {G.}~\bibnamefont {Frye}},\ }\bibfield  {title} {\bibinfo {title} {{Comment on the Single Particle Distribution in the Hydrodynamic and Statistical Thermodynamic Models of Multiparticle Production}},\ }\href {https://doi.org/10.1103/PhysRevD.10.186} {\bibfield  {journal} {\bibinfo  {journal} {Phys. Rev. D}\ }\textbf {\bibinfo {volume} {10}},\ \bibinfo {pages} {186} (\bibinfo {year} {1974})}\BibitemShut {NoStop}%
\bibitem [{\citenamefont {Bass}\ \emph {et~al.}(1998)\citenamefont {Bass} \emph {et~al.}}]{Bass:1998ca}%
  \BibitemOpen
  \bibfield  {author} {\bibinfo {author} {\bibfnamefont {S.~A.}\ \bibnamefont {Bass}} \emph {et~al.},\ }\bibfield  {title} {\bibinfo {title} {{Microscopic models for ultrarelativistic heavy ion collisions}},\ }\href {https://doi.org/10.1016/S0146-6410(98)00058-1} {\bibfield  {journal} {\bibinfo  {journal} {Prog. Part. Nucl. Phys.}\ }\textbf {\bibinfo {volume} {41}},\ \bibinfo {pages} {255} (\bibinfo {year} {1998})},\ \Eprint {https://arxiv.org/abs/nucl-th/9803035} {arXiv:nucl-th/9803035} \BibitemShut {NoStop}%
\bibitem [{\citenamefont {Bleicher}\ \emph {et~al.}(1999)\citenamefont {Bleicher} \emph {et~al.}}]{Bleicher:1999xi}%
  \BibitemOpen
  \bibfield  {author} {\bibinfo {author} {\bibfnamefont {M.}~\bibnamefont {Bleicher}} \emph {et~al.},\ }\bibfield  {title} {\bibinfo {title} {{Relativistic hadron hadron collisions in the ultrarelativistic quantum molecular dynamics model}},\ }\href {https://doi.org/10.1088/0954-3899/25/9/308} {\bibfield  {journal} {\bibinfo  {journal} {J. Phys. G}\ }\textbf {\bibinfo {volume} {25}},\ \bibinfo {pages} {1859} (\bibinfo {year} {1999})},\ \Eprint {https://arxiv.org/abs/hep-ph/9909407} {arXiv:hep-ph/9909407} \BibitemShut {NoStop}%
\bibitem [{\citenamefont {M{\"o}ller}\ \emph {et~al.}(2016)\citenamefont {M{\"o}ller}, \citenamefont {Sierk}, \citenamefont {Ichikawa},\ and\ \citenamefont {Sagawa}}]{Moller:2015fba}%
  \BibitemOpen
  \bibfield  {author} {\bibinfo {author} {\bibfnamefont {P.}~\bibnamefont {M{\"o}ller}}, \bibinfo {author} {\bibfnamefont {A.~J.}\ \bibnamefont {Sierk}}, \bibinfo {author} {\bibfnamefont {T.}~\bibnamefont {Ichikawa}},\ and\ \bibinfo {author} {\bibfnamefont {H.}~\bibnamefont {Sagawa}},\ }\bibfield  {title} {\bibinfo {title} {{Nuclear ground-state masses and deformations: FRDM(2012)}},\ }\href {https://doi.org/10.1016/j.adt.2015.10.002} {\bibfield  {journal} {\bibinfo  {journal} {Atom. Data Nucl. Data Tabl.}\ }\textbf {\bibinfo {volume} {109-110}},\ \bibinfo {pages} {1} (\bibinfo {year} {2016})},\ \Eprint {https://arxiv.org/abs/1508.06294} {arXiv:1508.06294 [nucl-th]} \BibitemShut {NoStop}%
\bibitem [{\citenamefont {Bilandzic}\ \emph {et~al.}(2011)\citenamefont {Bilandzic}, \citenamefont {Snellings},\ and\ \citenamefont {Voloshin}}]{Bilandzic:2010jr}%
  \BibitemOpen
  \bibfield  {author} {\bibinfo {author} {\bibfnamefont {A.}~\bibnamefont {Bilandzic}}, \bibinfo {author} {\bibfnamefont {R.}~\bibnamefont {Snellings}},\ and\ \bibinfo {author} {\bibfnamefont {S.}~\bibnamefont {Voloshin}},\ }\bibfield  {title} {\bibinfo {title} {{Flow analysis with cumulants: Direct calculations}},\ }\href {https://doi.org/10.1103/PhysRevC.83.044913} {\bibfield  {journal} {\bibinfo  {journal} {Phys. Rev. C}\ }\textbf {\bibinfo {volume} {83}},\ \bibinfo {pages} {044913} (\bibinfo {year} {2011})},\ \Eprint {https://arxiv.org/abs/1010.0233} {arXiv:1010.0233 [nucl-ex]} \BibitemShut {NoStop}%
\bibitem [{\citenamefont {Bilandzic}\ \emph {et~al.}(2014)\citenamefont {Bilandzic}, \citenamefont {Christensen}, \citenamefont {Gulbrandsen}, \citenamefont {Hansen},\ and\ \citenamefont {Zhou}}]{Bilandzic:2013kga}%
  \BibitemOpen
  \bibfield  {author} {\bibinfo {author} {\bibfnamefont {A.}~\bibnamefont {Bilandzic}}, \bibinfo {author} {\bibfnamefont {C.~H.}\ \bibnamefont {Christensen}}, \bibinfo {author} {\bibfnamefont {K.}~\bibnamefont {Gulbrandsen}}, \bibinfo {author} {\bibfnamefont {A.}~\bibnamefont {Hansen}},\ and\ \bibinfo {author} {\bibfnamefont {Y.}~\bibnamefont {Zhou}},\ }\bibfield  {title} {\bibinfo {title} {{Generic framework for anisotropic flow analyses with multiparticle azimuthal correlations}},\ }\href {https://doi.org/10.1103/PhysRevC.89.064904} {\bibfield  {journal} {\bibinfo  {journal} {Phys. Rev. C}\ }\textbf {\bibinfo {volume} {89}},\ \bibinfo {pages} {064904} (\bibinfo {year} {2014})},\ \Eprint {https://arxiv.org/abs/1312.3572} {arXiv:1312.3572 [nucl-ex]} \BibitemShut {NoStop}%
\bibitem [{\citenamefont {Bozek}\ and\ \citenamefont {Mehrabpour}(2020)}]{Bozek:2020drh}%
  \BibitemOpen
  \bibfield  {author} {\bibinfo {author} {\bibfnamefont {P.}~\bibnamefont {Bozek}}\ and\ \bibinfo {author} {\bibfnamefont {H.}~\bibnamefont {Mehrabpour}},\ }\bibfield  {title} {\bibinfo {title} {{Correlation coefficient between harmonic flow and transverse momentum in heavy-ion collisions}},\ }\href {https://doi.org/10.1103/PhysRevC.101.064902} {\bibfield  {journal} {\bibinfo  {journal} {Phys. Rev. C}\ }\textbf {\bibinfo {volume} {101}},\ \bibinfo {pages} {064902} (\bibinfo {year} {2020})},\ \Eprint {https://arxiv.org/abs/2002.08832} {arXiv:2002.08832 [nucl-th]} \BibitemShut {NoStop}%
\bibitem [{\citenamefont {{STAR Collaboration}}(2025)}]{STAR:2025elk}%
  \BibitemOpen
  \bibfield  {author} {\bibinfo {author} {\bibnamefont {{STAR Collaboration}}},\ }\bibfield  {title} {\bibinfo {title} {{Imaging nuclear shape through anisotropic and radial flow in high-energy heavy-ion collisions}},\ }\href {https://doi.org/10.1088/1361-6633/ae0fc3} {\bibfield  {journal} {\bibinfo  {journal} {Rept. Prog. Phys.}\ }\textbf {\bibinfo {volume} {88}},\ \bibinfo {pages} {108601} (\bibinfo {year} {2025})},\ \Eprint {https://arxiv.org/abs/2506.17785} {arXiv:2506.17785 [nucl-ex]} \BibitemShut {NoStop}%
\bibitem [{\citenamefont {Abdallah}\ \emph {et~al.}(2022{\natexlab{a}})\citenamefont {Abdallah} \emph {et~al.}}]{STAR:2021mii}%
  \BibitemOpen
  \bibfield  {author} {\bibinfo {author} {\bibfnamefont {M.}~\bibnamefont {Abdallah}} \emph {et~al.} (\bibinfo {collaboration} {STAR}),\ }\bibfield  {title} {\bibinfo {title} {{Search for the chiral magnetic effect with isobar collisions at $\sqrt {s_{NN}}$=200 GeV by the STAR Collaboration at the BNL Relativistic Heavy Ion Collider}},\ }\href {https://doi.org/10.1103/PhysRevC.105.014901} {\bibfield  {journal} {\bibinfo  {journal} {Phys. Rev. C}\ }\textbf {\bibinfo {volume} {105}},\ \bibinfo {pages} {014901} (\bibinfo {year} {2022}{\natexlab{a}})},\ \Eprint {https://arxiv.org/abs/2109.00131} {arXiv:2109.00131 [nucl-ex]} \BibitemShut {NoStop}%
\bibitem [{\citenamefont {Abdallah}\ \emph {et~al.}(2022{\natexlab{b}})\citenamefont {Abdallah} \emph {et~al.}}]{STAR:2022gki}%
  \BibitemOpen
  \bibfield  {author} {\bibinfo {author} {\bibfnamefont {M.}~\bibnamefont {Abdallah}} \emph {et~al.} (\bibinfo {collaboration} {STAR}),\ }\bibfield  {title} {\bibinfo {title} {{Collision-System and Beam-Energy Dependence of Anisotropic Flow Fluctuations}},\ }\href {https://doi.org/10.1103/PhysRevLett.129.252301} {\bibfield  {journal} {\bibinfo  {journal} {Phys. Rev. Lett.}\ }\textbf {\bibinfo {volume} {129}},\ \bibinfo {pages} {252301} (\bibinfo {year} {2022}{\natexlab{b}})},\ \Eprint {https://arxiv.org/abs/2201.10365} {arXiv:2201.10365 [nucl-ex]} \BibitemShut {NoStop}%
\bibitem [{\citenamefont {Mehrabpour}(2026)}]{Mehrabpour:2025ogw}%
  \BibitemOpen
  \bibfield  {author} {\bibinfo {author} {\bibfnamefont {H.}~\bibnamefont {Mehrabpour}},\ }\bibfield  {title} {\bibinfo {title} {{Imprint of {\ensuremath{\alpha}} clustering on ab initio correlations in relativistic light ion collisions}},\ }\href {https://doi.org/10.1103/cgcy-j883} {\bibfield  {journal} {\bibinfo  {journal} {Phys. Rev. C}\ }\textbf {\bibinfo {volume} {113}},\ \bibinfo {pages} {034909} (\bibinfo {year} {2026})},\ \Eprint {https://arxiv.org/abs/2506.12673} {arXiv:2506.12673 [nucl-th]} \BibitemShut {NoStop}%
\bibitem [{\citenamefont {Li}\ \emph {et~al.}(2026)\citenamefont {Li}, \citenamefont {Zhou},\ and\ \citenamefont {Ma}}]{Li:2025hae}%
  \BibitemOpen
  \bibfield  {author} {\bibinfo {author} {\bibfnamefont {P.}~\bibnamefont {Li}}, \bibinfo {author} {\bibfnamefont {B.}~\bibnamefont {Zhou}},\ and\ \bibinfo {author} {\bibfnamefont {G.-L.}\ \bibnamefont {Ma}},\ }\bibfield  {title} {\bibinfo {title} {{Identifying {\ensuremath{\alpha}}-Cluster Configurations in Ne20 via Ultracentral Ne+Ne Collisions}},\ }\href {https://doi.org/10.1103/tffz-8q1m} {\bibfield  {journal} {\bibinfo  {journal} {Phys. Rev. Lett.}\ }\textbf {\bibinfo {volume} {136}},\ \bibinfo {pages} {082302} (\bibinfo {year} {2026})},\ \Eprint {https://arxiv.org/abs/2504.04688} {arXiv:2504.04688 [nucl-th]} \BibitemShut {NoStop}%
\bibitem [{\citenamefont {Menon Kavumpadikkal~Radhakrishnan}\ \emph {et~al.}(2026)\citenamefont {Menon Kavumpadikkal~Radhakrishnan}, \citenamefont {Prasad}, \citenamefont {Mallick}, \citenamefont {Sahoo},\ and\ \citenamefont {Barnaf{\"o}ldi}}]{MenonKavumpadikkalRadhakrishnan:2026tfn}%
  \BibitemOpen
  \bibfield  {author} {\bibinfo {author} {\bibfnamefont {A.}~\bibnamefont {Menon Kavumpadikkal~Radhakrishnan}}, \bibinfo {author} {\bibfnamefont {S.}~\bibnamefont {Prasad}}, \bibinfo {author} {\bibfnamefont {N.}~\bibnamefont {Mallick}}, \bibinfo {author} {\bibfnamefont {R.}~\bibnamefont {Sahoo}},\ and\ \bibinfo {author} {\bibfnamefont {G.~G.}\ \bibnamefont {Barnaf{\"o}ldi}},\ }\href@noop {} {\bibinfo {title} {{Constraining {$\alpha$}-cluster compactness in {$^{16}\mathrm{O}$} and {$^{20}\mathrm{Ne}$} at {TeV} energies using azimuthal anisotropy}}} (\bibinfo {year} {2026}),\ \Eprint {https://arxiv.org/abs/2607.26758} {arXiv:2607.26758 [hep-ph]} \BibitemShut {NoStop}%
\bibitem [{\citenamefont {Wang}\ \emph {et~al.}(2026)\citenamefont {Wang}, \citenamefont {Chen}, \citenamefont {Jia}, \citenamefont {Ma},\ and\ \citenamefont {Zhang}}]{Wang:2024mce}%
  \BibitemOpen
  \bibfield  {author} {\bibinfo {author} {\bibfnamefont {Z.}~\bibnamefont {Wang}}, \bibinfo {author} {\bibfnamefont {J.}~\bibnamefont {Chen}}, \bibinfo {author} {\bibfnamefont {J.}~\bibnamefont {Jia}}, \bibinfo {author} {\bibfnamefont {Y.-G.}\ \bibnamefont {Ma}},\ and\ \bibinfo {author} {\bibfnamefont {C.}~\bibnamefont {Zhang}},\ }\bibfield  {title} {\bibinfo {title} {{Disentangling nuclear structure through multiparticle azimuthal correlations in high-energy isobar collisions}},\ }\href {https://doi.org/10.1103/47jc-xzxm} {\bibfield  {journal} {\bibinfo  {journal} {Phys. Rev. C}\ }\textbf {\bibinfo {volume} {113}},\ \bibinfo {pages} {034905} (\bibinfo {year} {2026})},\ \Eprint {https://arxiv.org/abs/2409.15040} {arXiv:2409.15040 [nucl-th]} \BibitemShut {NoStop}%
\bibitem [{\citenamefont {Magdy}(2024)}]{Magdy:2024ooh}%
  \BibitemOpen
  \bibfield  {author} {\bibinfo {author} {\bibfnamefont {N.}~\bibnamefont {Magdy}},\ }\bibfield  {title} {\bibinfo {title} {{Characterizing the initial- and final-state effects in isobaric collisions at energies available at the BNL Relativistic Heavy Ion Collider}},\ }\href {https://doi.org/10.1103/PhysRevC.109.024906} {\bibfield  {journal} {\bibinfo  {journal} {Phys. Rev. C}\ }\textbf {\bibinfo {volume} {109}},\ \bibinfo {pages} {024906} (\bibinfo {year} {2024})},\ \Eprint {https://arxiv.org/abs/2401.04083} {arXiv:2401.04083 [nucl-th]} \BibitemShut {NoStop}%
\bibitem [{\citenamefont {Magdy}(2023)}]{Magdy:2022cvt}%
  \BibitemOpen
  \bibfield  {author} {\bibinfo {author} {\bibfnamefont {N.}~\bibnamefont {Magdy}},\ }\bibfield  {title} {\bibinfo {title} {{Impact of nuclear deformation on collective flow observables in relativistic U+U collisions}},\ }\href {https://doi.org/10.1140/epja/s10050-023-00982-0} {\bibfield  {journal} {\bibinfo  {journal} {Eur. Phys. J. A}\ }\textbf {\bibinfo {volume} {59}},\ \bibinfo {pages} {64} (\bibinfo {year} {2023})},\ \Eprint {https://arxiv.org/abs/2206.05332} {arXiv:2206.05332 [nucl-th]} \BibitemShut {NoStop}%
\bibitem [{\citenamefont {Aad}\ \emph {et~al.}(2019)\citenamefont {Aad} \emph {et~al.}}]{ATLAS:2019pvn}%
  \BibitemOpen
  \bibfield  {author} {\bibinfo {author} {\bibfnamefont {G.}~\bibnamefont {Aad}} \emph {et~al.} (\bibinfo {collaboration} {ATLAS}),\ }\bibfield  {title} {\bibinfo {title} {{Measurement of flow harmonics correlations with mean transverse momentum in lead-lead and proton-lead collisions at $\sqrt{s_{NN}}=5.02$ TeV with the ATLAS detector}},\ }\href {https://doi.org/10.1140/epjc/s10052-019-7489-6} {\bibfield  {journal} {\bibinfo  {journal} {Eur. Phys. J. C}\ }\textbf {\bibinfo {volume} {79}},\ \bibinfo {pages} {985} (\bibinfo {year} {2019})},\ \Eprint {https://arxiv.org/abs/1907.05176} {arXiv:1907.05176 [nucl-ex]} \BibitemShut {NoStop}%
\bibitem [{\citenamefont {Giacalone}\ \emph {et~al.}(2020)\citenamefont {Giacalone}, \citenamefont {Schenke},\ and\ \citenamefont {Shen}}]{Giacalone:2020byk}%
  \BibitemOpen
  \bibfield  {author} {\bibinfo {author} {\bibfnamefont {G.}~\bibnamefont {Giacalone}}, \bibinfo {author} {\bibfnamefont {B.}~\bibnamefont {Schenke}},\ and\ \bibinfo {author} {\bibfnamefont {C.}~\bibnamefont {Shen}},\ }\bibfield  {title} {\bibinfo {title} {{Observable signatures of initial state momentum anisotropies in nuclear collisions}},\ }\href {https://doi.org/10.1103/PhysRevLett.125.192301} {\bibfield  {journal} {\bibinfo  {journal} {Phys. Rev. Lett.}\ }\textbf {\bibinfo {volume} {125}},\ \bibinfo {pages} {192301} (\bibinfo {year} {2020})},\ \Eprint {https://arxiv.org/abs/2006.15721} {arXiv:2006.15721 [nucl-th]} \BibitemShut {NoStop}%
\end{thebibliography}%

\end{document}